\documentclass[11pt,a4paper]{article}
\usepackage{amsmath,amssymb}

\newcommand{\fracd}[2]{{\displaystyle\frac{#1}{#2}}}
\newcommand{\fract}[2]{{\frac{#1}{#2}}}
\newcommand{\fracs}[2]{{\scriptstyle\frac{#1}{#2}}}

\newcommand{\za} {{\alpha}}   
\newcommand{\zb} {{\beta}}    
\newcommand{\zm} {{\mu}}      
\newcommand{\zn} {{\nu}}      

\newcommand{\z}{{\phantom{\za}}}

\newcommand{\ZA} {{\scriptscriptstyle{\!A}}}   
\newcommand{\ZB} {{\scriptscriptstyle{\!B}}}   
\newcommand{\ZC} {{\scriptscriptstyle{\!C}}}   

\newcommand{\Z}{{\phantom{\scriptscriptstyle{\ZA}}}}

  \newcommand{\brWF}{{\sigma}}    

  \newcommand{\B}{{\boldsymbol B}}
  \newcommand{\br}{{\mathbf{b}}}          

  \newcommand{\FB}{{{F}}}          
  \newcommand{\FBdev}{{V}}       
  \newcommand{\PB}{{{P}}}          

  \newcommand{\Fbr}{{\varkappa}}          
  \newcommand{\Pbr}{{p}}           

  \newcommand{\rt} {{\ensuremath{\mathchoice%
       {\sqrt{\MB^2{-}\brBox} \,}               
       {\sqrt{\MB^2{-}\brBox }\,}               
       {\sqrt{\MB^{\!2}{-}\brBox }\,}     
       {\sqrt{\MB^{\!2}{-}\brBox \,}} }}} 

  \newcommand{\Feff} {{\mbox{\boldmath$F$}}^{\,\mathrm{brane}}}

  \newcommand{\Feffo} {{\ensuremath{\mathchoice%
       {{\textstyle{\boldsymbol F}_{\!0}}}               
       {{\textstyle{\boldsymbol F}_{\!0}}}               
       {{\scriptstyle{\boldsymbol F}_{\!0}^{}}}          
       {{\scriptscriptstyle{\boldsymbol F}_{\!0}}} }}}   

  \newcommand{\MB}{{M}}      
  \newcommand{\Mbr}{{\mu}}             
  \newcommand{\Mdgp}{m}       

  \newcommand{\Mp}{{m_{\!_+}\!}}    
  \newcommand{\Mm}{{m_{\!_-}\!}}    
  \newcommand{\Mpm}{m_{_{\pm}}\!}   
  \newcommand{\Mx}{m}               

  \newcommand{\hdM}{\vartriangle}   

  \newcommand{\brBox}{{\square}}                 

  \newcommand{\HU}{{U}}
  \newcommand{\HW}{{W}}

  \newcommand{\MW}{{w}}

  \newcommand{\MC}{{C}}                    

  \newcommand{\Abr}{{\mathcal{A}}}

  \newcommand{\Fu}{{{u}}}
  \newcommand{\Fv}{{{v}}}

  \newcommand{\RB} {{^{\scriptscriptstyle\boldsymbol B}\!{R}}}     
  \newcommand{\Rbr} {{^{\scriptscriptstyle\br}\!R}}           

  \newcommand{\CMdim}[1] {{\mathfrak{M}^{#1}}}     

  \newcommand{\ad}{{\mathrm{ad}}} 

  \newcommand{\pt}{s}          

  \newcommand{\brint}{\int\limits_{\br}\!d^{d}\!x\,\sqrt{g}\;}        

  \newcommand{\GammaF}[1] {{\ensuremath{\mathchoice%
    {\,{\varGamma}{\!\textstyle\left(#1\right)}}
    {\,{\varGamma}{\!\textstyle\big(#1\big)}}
    {\,{\varGamma}{\!\scriptstyle(#1)}}
    {\,{\varGamma}{\!\scriptscriptstyle\left(#1\right)}} }}}

  \newcommand{\erf} {\mathop{\mathrm{erf}}\nolimits} 
  \DeclareMathOperator{\erfc} {erfc}  

  \newcommand{\HGF}[3]    {\mathop{_{_2}\!
  {F}_{_1}}\!\left({#1;\,#2;\,#3}\right)}

  \newcommand{\ResHGFp}[2]    {\mathop{\Phi}\!_{_{#1}}\!\big({\textstyle #2}\big)}  %
\begin{document}

\title{\bf The Schwinger-DeWitt technique for quantum effective
action in brane induced gravity models}
\author{
A.~O.~Barvinsky\footnote{{\bf e-mail}: barvin@lpi.ru}\,,\;
D.~V.~Nesterov\footnote{{\bf e-mail}: nesterov@lpi.ru}
\\
\small{\em Theory Department, Lebedev Physics Institute,} \\
\small{\em Leninsky prospect 53, Moscow, Russia, 119991}
}
\date{}
\maketitle

\begin{abstract}
We develop the Schwinger-DeWitt technique for the covariant
curvature expansion of the quantum effective action for brane
induced gravity models in curved spacetime. This expansion has a
part nonanalytic in DGP type scale parameter $\Mdgp$, leading to the
cutoff scale which is given by the geometric average of the mass of
the quantum field in the bulk $\MB$ and $\Mdgp$. This cutoff $M_{\rm
cutoff}=\sqrt{\MB\Mdgp}$ is much higher than the analogous strong
coupling scale of the DGP model treated by weak field expansion in
the tree-level approximation. The lowest orders of this curvature
expansion are calculated for the case of the scalar field in the
$(d+1)$-dimensional bulk with the brane carrying the $d$-dimensional
kinetic term of this field. The ultraviolet divergences in this
model are obtained for a particular case of $d=4$.
\end{abstract}

\section{Introduction}
  \hspace{\parindent}
Modified theories of gravity in the form of braneworld models can in
principle account for the phenomenon of dark energy
\cite{DGP,Deffayet} as well as for nontrivial compactifications of
multi-dimensional string models. It becomes increasingly more
obvious that one should include in such models the analysis of
quantum effects beyond the tree-level approximation
\cite{quantumDGP}. This is the only way to reach an ultimate
conclusion on the resolution of such problems as the presence of
ghosts \cite{ghosts} and low strong-coupling scale
\cite{scale0,scale,scale2}. Quantum effects in brane models are also
important for the stabilization of extra dimensions
\cite{GarPujTan}, fixing the cross-over scale in the Brans-Dicke
modification of the DGP model \cite{Pujolas} and in the recently
suggested mechanism of the cosmological acceleration generated by
the four-dimensional conformal anomaly \cite{slih}.

A general framework for treating quantum effective actions in brane
models (or, more generally, models with timelike and spacelike
boundaries) was recently suggested in
\cite{BKRK,gospel,qeastb,toyDGP}. The main peculiarity of these
models is that due to quantum field fluctuations on the branes the
field propagator is subject to generalized Neumann boundary
conditions involving normal and tangential derivatives on the
brane/boundary surfaces. This presents both technical and conceptual
difficulties, because such boundary conditions are much harder to
handle than the simple Dirichlet ones. The method of \cite{qeastb}
provides a systematic reduction of the generalized Neumann boundary
conditions to Dirichlet conditions. As a byproduct it disentangles
from the quantum effective action the contribution of the surface
modes mediating the brane-to-brane propagation, which play a very
important role in the zero-mode localization mechanism of the
Randall-Sundrum type \cite{RS}. The purpose of this work is to make
the next step --- to extend a well-known Schwinger-DeWitt technique
\cite{DeWitt,PhysRep,McKean-Singer,Vassilevich} to the calculation
of this contribution in the DGP model in a weakly curved spacetime
in the form of the {\em covariant} curvature expansion.

Briefly the method of \cite{qeastb} looks as follows. The action of
a (free field) brane model generally contains the bulk and the brane
parts,
    \begin{eqnarray}
     S[\,\varPhi\,]=\frac12\int\limits_{\rm\bf B} d^{d+1}\!X\sqrt{G}\,
       \; \varPhi(X)\,{\FB}(\nabla_{\!X})\,\varPhi(X)
      +\frac12 \brint
       \; \varphi(x)\,
       \Fbr(\nabla_{\!x})\,\varphi(x) \, ,   \label{QuadraticAction}
    \end{eqnarray}
where the $(d\!+\!1)$-dimensional bulk and the $d$-dimensional brane
coordinates are labeled respectively by $X=X^A$ and $x=x^\mu$, and
the boundary values of bulk fields $\varPhi(X)$ on the brane/boundary
${\rm\bf b}=\partial\rm\bf B$ are denoted by $\varphi(x)$,
    \begin{eqnarray}
     \varPhi(X)\,\Big|_{\,\rm\bf b}=\varphi(x),        \label{2}
    \end{eqnarray}
$G$ and $g$ are the determinants of the bulk $G_{AB}$ and
$g_{\mu\nu}$ metrics respectively.

The kernel of the bulk Lagrangian is given by the second order
differential operator $\FB(\nabla_X)$, whose covariant derivatives
$\nabla_X$ in (\ref{QuadraticAction}) are integrated by parts in
such a way that they form bilinear combinations of first order
derivatives acting on two different fields. Integration by parts in
the bulk gives nontrivial surface terms on the brane/boundary. In
particular, this operation results in the Wronskian relation for
generic test functions $\phi_{1,\,2}(X)$,
    \begin{eqnarray}
     \int\limits_{\rm\bf B} d^{\,d{+}1}X\,\sqrt{G}
        \left(\,\varPhi_1\overrightarrow{\FB}\!(\nabla_{\!X})\,\varPhi_2-
              \varPhi_1\!\overleftarrow{\FB}\!(\nabla_{\!X})\,\varPhi_2\right)=
     -\brint     
        \left(\,\varPhi_1 \overrightarrow{W} \varPhi_2-
              \varPhi_1 \overleftarrow{W} \varPhi_2\right).
    \label{WronskianRelation}
    \end{eqnarray}         
Arrows everywhere here indicate the direction of action of
derivatives either on $\varPhi_1$ or $\varPhi_2$.

The brane part of the action contains as a kernel some local
operator $\hat\Fbr(\nabla)$, $\nabla\equiv\nabla_x$. Its order in
derivatives depends on the model in question. In the Randall-Sundrum
model \cite{RS}, for example, it is for certain gauges just an
ultralocal multiplication operator generated by the tension term on
the brane. In the Dvali-Gabadadze-Porrati (DGP) model \cite{DGP}
this is a second order operator induced by the brane Einstein term
on the brane, $\hat\Fbr(\nabla)\sim\nabla\nabla/\Mdgp$, where
$\Mdgp$ is the DGP scale which is of the order of magnitude of the
horizon scale, being responsible for the cosmological acceleration
\cite{Deffayet}. In the context of the Born-Infeld action in D-brane
string theory with vector gauge fields, $\Fbr(\nabla)$ is a
first-order operator \cite{open}.

In all these cases the variational procedure for the action
(\ref{QuadraticAction}) with dynamical (not fixed) fields on the boundary
$\varphi(x)$ naturally leads to generalized Neumann boundary
conditions of the form
    \begin{eqnarray}
     \left.\Big(\overrightarrow{W}(\nabla_{\!X})
     +\Fbr(\nabla)\Big)\,\varPhi\,\right|_{\,\rm\bf b}
     =0,
    \label{BoundaryCondition}
    \end{eqnarray}
which uniquely specify the propagator of quantum fields and,
therefore, a complete Feynman diagrammatic technique for the system
in question. The method of \cite{qeastb} allows one to
systematically reduce this diagrammatic technique to the one subject
to the Dirichlet boundary conditions $\varPhi|_{\,\rm\bf b}=0$. The
main additional ingredient of this reduction procedure is the brane
operator $\Feff(x,x')$ which is
constructed from the Dirichlet Green's function $G_D(X,X')$ of the
operator $F(\nabla)$ in the bulk,
    \begin{eqnarray}
    \Feff(x,x')=-
    \overrightarrow{W}(\nabla_{\!X}\!)\,G_{D}(X,X')\,
    \overleftarrow{W}(\nabla_{\!X'})
    \,\Big|_{\,X=e(x),\,X'=e(x')}
    +\Fbr(\nabla)\,\delta(x,x')\ .           \label{5}
    \end{eqnarray}
This expression expresses the fact that the kernel of the Dirichlet
Green's function is being acted upon both arguments by the Wronskian
operators with a subsequent restriction to the brane, with $X=e(x)$
denoting the brane embedding function.

As shown in \cite{qeastb}, this operator determines the
brane-to-brane propagation of the physical modes in the system with
the classical action (\ref{QuadraticAction}) (its inverse is the brane-to-brane
propagator) and additively contributes to its full one-loop
effective action according to
    \begin{eqnarray}
     \mbox{\boldmath$\varGamma$}_{\rm 1-loop}
     \equiv \frac12\; {\rm Tr}_N^{(d+1)}
     \ln \FB=\frac12\;
     {\rm Tr}_D^{(d+1)}\ln \FB
     + \frac12\;{\rm Tr}^{(d)}
     \ln \Feff,                 \label{Neumann-to-Dirichlet_Reduction}
    \end{eqnarray}
where ${\rm Tr}_{D,N}^{(d+1)}$ denotes functional traces of the bulk
theory subject to Dirichlet and Neumann boundary conditions,
respectively, while ${\rm Tr}^{(d)}$ is a functional trace in the
boundary $d$-dimensional theory. The full quantum effective action
of this model is obviously given by the functional determinant of
the operator $F(\nabla_X)$ subject to the generalized Neumann
boundary conditions (\ref{5}), and the above equation reduces its
calculation to that of the Dirichlet boundary conditions plus the
contribution of the brane-to-brane propagation.

Here we apply (\ref{Neumann-to-Dirichlet_Reduction}) to a simple
model of a scalar field which mimics in particular the properties of
the brane-induced gravity models and the DGP model \cite{DGP}. This
is the $(d\!+\!1)$-dimensional massive scalar field
$\varPhi(X)=\varPhi(x,y)$ with mass $M$ living in the {\em curved}
half-space $y\geq 0$ with the additional $d$-dimensional kinetic
term for $\varphi(x)\equiv\varPhi(x,0)$ localized at the brane
(boundary) at $y=0$,
    \begin{eqnarray}
    &&S[\,\phi\,]=\frac12\int\limits_{y\geq 0}
    d^{d+1}X\,\sqrt{G}\,\Big((\nabla_X\varPhi(X))^2
    +\MB^2\varPhi^2(X)+\PB(X)\varPhi^2(X)\Big)
    \nonumber\\
    &&\qquad\qquad+\frac1{4\Mdgp}\int
    d^dx \,\sqrt{g}\,
    \Big(\big(\nabla_x\varphi(x)\big)^2
    +\Mbr^2\varphi^2(x)
    +\Pbr(x)\varphi^2(x)\Big).           \label{1.1}
    \end{eqnarray}
Here and in what follows we work in a Euclidean (positive-signature)
spacetime. Therefore, this action corresponds to the following
choice of $\FB(\nabla_X)$ in terms of $(d\!+\!1)$-dimensional
covariant d'Alembertian (Laplacians)
    \begin{eqnarray}
    &&\FB(\nabla_X)=-\Box^{(\,d+1)}+\MB^2+\PB=
    -G^{AB}\nabla_A\nabla_B +\MB^2+\PB.              \label{6}
    \end{eqnarray}
In the normal Gaussian coordinates its Wronskian operator is given
by $W=-\partial_y$ --- the normal derivative with respect to
outward-pointing normal to the brane,  and the boundary operator
$\Fbr(\nabla)$ equals in terms of the $d$-dimensional d'Alembertian
    \begin{eqnarray}
    \Fbr(\nabla)=\frac1{2\Mdgp}\,
    \big(-\Box +\Mbr^2+\Pbr\,\big),\, \,\,\,\,
    \Box=\Box^{(d)}\equiv
    g^{\mu\nu}\nabla_\mu\nabla_\nu,              \label{hFbr_def}
    \end{eqnarray}
where the dimensional parameter $\Mdgp$ mimics the role of the DGP
scale \cite{DGP}. Thus, the generalized Neumann boundary conditions
in this model involve second-order derivatives tangential to the
brane,
    \begin{eqnarray}
    \left(\partial_y
    -\frac{-\Box +\Mbr^2+\Pbr}{2\Mdgp}\right)\,\varPhi(X)\,
    \Big|_{\,\rm\bf b}=0,               \label{1.3}
    \end{eqnarray}
cf. (\ref{BoundaryCondition}) with $W=-\partial_y$ and $\varkappa$
given by (\ref{hFbr_def}).

As was shown \cite{toyDGP}, the flat space brane-to-brane operator for
such a model without potential terms has the form of the pseudodifferential operator with
the flat-space $\Box$,
    \begin{eqnarray}
    \Feff(\nabla)
    =\frac1{2\Mdgp}\,(-\Box+\Mbr^2+2\Mdgp\rt).          \label{9}
    \end{eqnarray}
In the massless case of the DGP model \cite{DGP}, $\MB=0$, this
operator is known to mediate the gravitational interaction on the
brane, interpolating between the four-dimensional Newtonian law at
intermediate distances and the five-dimensional law at the horizon
scale $\sim 1/\Mdgp$ \cite{scale}.

Here we generalize this construction to a curved spacetime and
expand the brane-to-brane operator and its effective action in
covariant curvature series. This is the expansion in powers of the
bulk curvature $\RB$, extrinsic curvature of the brane
$k_{\alpha\beta}$, the potential terms of the bulk $\PB$ and brane
$\Pbr$ operators and their covariant derivatives
--- all taken at the location of the brane. The expansion starts
with the approximation (\ref{9}) based on the {\em full covariant}
d'Alembertian on the brane. We present a systematic technique of
calculating curvature corrections in
(\ref{Neumann-to-Dirichlet_Reduction}) and rewrite their nonlocal
operator coefficients --- functions of the covariant $\brBox$ -- in
the form of the generalized (weighted) proper time representation.


The success of the conventional Schwinger-DeWitt method is based on
the fact that the one-loop effective action of the operator, say
${-}\Box+M^2$, has a proper time representation
    \begin{eqnarray}
     &&\frac12\,{\rm Tr}\,\ln\,\Big({-}\Box + M^2\Big)
      = -\frac12 \int\limits_0^\infty
      \frac{ds}s\;e^{-s\,M^2}\,
      {\rm Tr}\:e^{s\,\Box}\;.             \label{proptime}
    \end{eqnarray}
In view of the well-known small time expansion for the heat kernel
\cite{DeWitt,PhysRep},
    \begin{eqnarray}
    e^{\,s\,\Box}\delta(x,x')=\frac1{(4\pi s)^{d/2}}
    D^{1/2}(x,x')\,
    e^{-\sigma(x,x')/2s}\sum_{n=0}^\infty\,s^n\,
    a_{n}(x,x'),                  \label{braneHK_Ansatz}
    \end{eqnarray}
($\sigma(x,x')$ is the geodetic world function, $D(x,x')$ is the
associated Van Vleck determinant and $a_{n}(x,x')$ are the
Schwinger-DeWitt or Gilkey-Seely coefficients) the curvature
expansion eventually reduces to the calculation of the coincidence
limits of $a_n(x,x')$ and a trivial proper time integration
resulting in the inverse mass expansion
    \begin{eqnarray}
    \frac12\,\mathrm{Tr}\,\ln\,\Big(M^2-\Box\Big)
    =-\frac12 \frac{M^d}{(4\pi)^{d/2}}
    \sum\limits_{n=0}^\infty
    \frac{\GammaF{n+{d/2}}}{M^{2n}}\,
    \int\!\!  dx\,\sqrt{g}\;\, a_{n}(x,x). \label{SchwDW}
    \end{eqnarray}

As we will show below, the calculation of the brane effective action
differs from the conventional Schwinger-DeWitt case in that the
proper time integral (\ref{proptime}) contains in the integrand a
certain extra weight function $w(s)$, and instead of just ${\rm
Tr}\,e^{s\Box}$ one has to calculate the trace of the heat kernel
acted upon by a certain local differential operator ${\rm
Tr}\,\big(W(\nabla)e^{s\Box}\big)$. This again reduces to the
calculation of the coincidence limits --- this time of the multiple
covariant derivatives of $a_{n}(x,x')$, $\sigma(x,x')$ and $D(x,x')$
--- the task easily doable within a conventional DeWitt recurrence
procedure.

The result of this calculation is peculiar. Unlike the usual
Schwinger-DeWitt expansion (\ref{SchwDW}) the brane effective action
takes the form
    \begin{eqnarray}
     &&\frac12\;\mathrm{Tr}\ln{\Feff}
      =\left(\frac{M
      m}{4\pi}\right)^{d/2}\sum\limits_{N=0}^\infty
      \frac1{M^N}\;\sum\limits_{\,i\leq N\,}
     \frac{O(\CMdim{2N-i})}{\Mdgp^{N-i}}\nonumber\\
     &&\qquad\qquad\qquad\quad
     +\frac{M^d}{(4\pi)^{d/2}}
      \sum\limits_{N=0}^\infty
      \frac1{M^N}\sum\limits_{i\leq N}
     \Mdgp^i\;
     O(\CMdim{N-i})\,,      \label{efficiency0}
    \end{eqnarray}
where $O(\CMdim{k})$ represent the integrals over the brane/boundary
space of local invariants of dimensionality $k$ in units of mass or
inverse length. With this notation, in particular,
$a_n(x,x)=O(\CMdim{2n})$. More generally, these invariants (or
spacetime covariant higher-dimensional operators) are composed of
the powers of the bulk and brane curvature, extrinsic curvature of
the brane/boundary, the potential terms of the bulk and brane
operators and their covariant derivatives.

The main difference of (\ref{efficiency0}) from (\ref{SchwDW}) is
that in addition to a usual part analytic in $\Mdgp$ with a typical
$\MB$-dependence (second series in (\ref{efficiency0})) we also have
the part singular in $\Mdgp\to 0$ with a qualitatively different
analytic dependence on the bulk mass ($\MB^{d/2-N}$ instead of
$\MB^{d-N}$). This property was recently discovered for the
effective potential in the toy model of the DGP type \cite{toyDGP}.
Physically this leads to an essential modification of the
perturbation theory cutoff --- the domain of validity of the local
expansion $\CMdim{}\ll M_{\rm cutoff}$. It reduces this cutoff from
$M_{\rm cutoff}=\MB$ to
    \begin{eqnarray}
    M_{\rm cutoff}=\sqrt{\MB\Mdgp}.    \label{cutoff}
    \end{eqnarray}
In physically interesting brane models with $\Mdgp\ll\MB$ this
implies essential reduction of $M_{\rm cutoff}$ and signifies the
problem of a low strong coupling scale \cite{scale}. While in
\cite{scale} this phenomenon was observed in the tree-level theory,
here we extend it to the quantum one-loop approximation.

As an application of this generalized Schwinger-DeWitt expansion we
calculate the one-loop brane effective action of the quantum scalar
field with the accuracy $O(\CMdim{2})$. In this approximation the
basis of local curvature invariants includes one structure as a
cosmological term, two structures linear in the extrinsic curvature
and the potential term of the brane operator (\ref{hFbr_def}) and
seven structures of dimensionality $(\CMdim{2})$,
    \begin{eqnarray}
     &&O(\CMdim{0})=\brint,                 \label{zero}\\
     &&O(\CMdim{1})=\brint k,\,\,\,
     \brint \frac{\Pbr}{2\Mdgp},             \label{one}\\
     &&O(\CMdim{2})=\brint \RB,\,\,\,
     \brint\RB_{nn},\,\,\,
     \brint k_{\alpha\beta}^2,\,\,\,\brint k^2,\nonumber\\
     &&\qquad\qquad\brint\PB,\,\,\,
     \brint \left(\frac{\Pbr}{2\Mdgp}
     \right)^2,\,\,\,
     \brint k \frac{\Pbr}{2\Mdgp}.    \label{two}
    \end{eqnarray}
Here $\RB$ is a bulk scalar curvature,
$\RB_{nn}=\RB_{\ZA\ZB}\,n^\ZA n^\ZB$ is the projection of
the bulk Ricci tensor on the normal vector $n^\ZA$ to the brane,
$k_{\alpha\beta}$ is the extrinsic curvature of the brane\footnote{In this article we use the following sign convention for extrinsic curvature
    \begin{eqnarray}
     k_{\za\zb}= e^\ZA_{(\za} e^\ZB_{\zb)} \big(\nabla_{\!_X}\big)_{\ZA} n_{\ZB}\,,
     \nonumber
    \end{eqnarray}
where $\{e^\ZA_\za\}$ is a holonomic basis tangent to brane, $n_\ZA$ -- unit inward normal vector to the brane.} and
$k=g^{\alpha\beta}k_{\alpha\beta}$ is its trace, and $\PB$ and
$\Pbr$ represent the bulk and brane potential terms of relevant
operators $F(\nabla_X)$ and $\Fbr(\nabla_x)$ introduced above. Below
we find explicit coefficients of these structures in
(\ref{efficiency0}) as nontrivial functions of mass parameters
$\MB$, $\Mdgp$ and $\Mbr$ and find UV divergences in this model.

\section{Perturbation theory for the bulk Green's function
and brane-to-brane inverse propagator.}
\hspace{\parindent}
In normal Gaussian coordinates the covariant bulk d'Alembertian
decomposes as $\Box^{\,(d+1)}_X=\partial_y^2+\Box_x(y)+...$, where
ellipses denote depending on spin terms at most linear in
derivatives\footnote{This term for a general spin structurally has
the form $k\nabla_X+k^2+(\nabla k)+R$ where $k$ is the extrinsic
curvature of $y={\rm const}$ slices and $R$ is the bulk curvature.}
and $\Box_x(y)$ is a covariant d'Alembertian on the slice of
constant coordinate $y$. Therefore the full bulk operator takes the
form
    \begin{eqnarray}
     &&\FB(\nabla)
     =-\Box^{(\,d+1)}_X+\MB^2+\PB(X)\nonumber\\
     &&\qquad\quad
     =-\partial_y^2-\Box+\MB^2-\FBdev(X\,|\,\partial_y,\nabla)
     \;\equiv\; \FB_0-\FBdev,\,\,\,\,\,\,\Box\equiv\Box_x(0),
     \label{FB_Splitting}
    \end{eqnarray}
in which all nontrivial $y$-dependence is isolated as a perturbation
term $V(X\,|\,\partial_y,\nabla)\equiv V(y,\partial_y)$ --- a
first-order differential operator in $y$, proportional to the
extrinsic and bulk curvatures, and of second order in brane
derivatives $\nabla$ which we do not explicitly indicate here by
assuming that they are encoded in the operator structure of
$V(y,\partial_y)$. In particular, it includes the difference
$\Box_x(0)-\Box(y)\equiv\Box-\Box(y)$ expandable in Taylor series in
$y$.

The kernel of the bulk Green's function can formally be written as a
$y$-dependent nonlocal operator acting on the $d$-dimensional brane
--- some non-polynomial function of the brane covariant derivative
    \begin{eqnarray}
     G_D(X,X')=G_D(y,y'|\,\nabla)\,\delta(x,x')\;.
    \end{eqnarray}
The perturbation expansion for $G_D(y,y'|\,\nabla)$ is usual
    \begin{eqnarray}
     G_D
     = G_D^0+G^0_D \FBdev G^0_D+...
     = G^0_D\sum_{n=0}^\infty \big(\FBdev\, G^0_D\big)^n\;,
    \label{Gpert}
    \end{eqnarray}
where $G_D^0$ is the propagator for operator $\FB_0$ obeying Dirichlet boundary conditions
and  the composition law includes the integration over the bulk
coordinates, like for example in the first subleading term
    \begin{eqnarray}
     G^0_D \FBdev G^0_D(y,y')=\int_0^\infty dy''\,
     G^0_D(y,y'') \FBdev(y'',\partial_{y''}) G^0_D(y'',y')\;.
    \end{eqnarray}

The lowest order Green's function in the half-space of the DGP model
setting --- the Green's function of $\FB_0=-\partial_y^2-\Box+\MB^2$
subject to Dirichlet conditions on the brane $y=0$ and at infinity
--- reads as follows
    \begin{eqnarray}
     &&G_{D}^0(y,y')=\frac{e^{-|\,y-y'|\,\rt}
     -e^{-(y+y')\,\rt}}{2\rt}\;.
    \end{eqnarray}
We want to stress that here we assume the exact (curved)
$d$-dimensional d'Alembertian $\Box$ depending on the induced metric
of the brane $g_{\mu\nu}(x)$. This means that in the lowest order
approximation the underlying spacetime is not flat, but rather has a
nontrivial but constant in $y$ metric of constant $y$ slices.
Correspondingly in the zeroth order we have
    \begin{eqnarray}
     -\big[\overrightarrow{W} G_{D}(y,y') \overleftarrow{W} \big]_{\,y=y'=0}^{\,0}
     \;=\; -\overrightarrow{\partial_y} G_{D}^0(y,y') \overleftarrow{\partial_y} \Big|_{\,y=y'=0}
     = \rt\;.
    \label{G0}
    \end{eqnarray}

The perturbation of the bulk operator can be expanded in Taylor
series in $y$, so that it reads
    \begin{eqnarray}
     && \FBdev=k(x,y)\partial_y + \square(y)-\square-\PB(x,y)
           \equiv \sum\limits_{k=0}^{\infty}\frac1{k!}\Fu_k\;y^k\;\partial_y
                - \sum\limits_{k=0}^{\infty}\frac1{k!}\Fv_k\;y^k\;,
     \label{FBdev_y_expansion}
    \end{eqnarray}
where $\Fu_k(\nabla)$ and $\Fv_k(\nabla)$ form a set of
$y$-independent {\em local $d$-dimensional covariant} operators of
maximum second order in $\nabla_x$. The coefficients of these
operators are given by the powers of the bulk and brane curvature,
the extrinsic curvature of the brane, the potential term $P$ and the
covariant derivatives of all these quantities --- all of them taken
at the brane.

Below we present them up to the first order in the bulk and brane
curvature and to the second order in the extrinsic curvature of the
brane. Working in gauss normal coordinate system for the case of a single quantum
scalar field they read as
    \begin{eqnarray}
     && \Fu_0=k(x,y)\big|_{\,y=0}\equiv k, \nonumber\\
     && \Fu_1=\partial_y k(x,y)\big|_{\,y=0}
       =-\RB_{nn} -k_{\zm\zn}^2
     \label{Fu_Structures}
    \end{eqnarray}
and
    \begin{eqnarray}
     && \Fv_0 = \PB(x,y)\Big|_{\,y=0}\equiv P, \nonumber\\
     && \Fv_1 = \partial_y \big(-\Box_x(y)+\PB(x,y)\big)\Big|_{\,y=0}
       =2k^{\za\zb}\,\nabla_{\za}\nabla_{\zb}
       + 2(\nabla_{\za}k^{\za\zb})\,\nabla_{\zb}
       - (\nabla^{\zb} k)\,\nabla_{\zb} + O[\CMdim{3}],    \nonumber\\
     && \Fv_2 = \partial_y^2\big(-\Box_x(y)+\PB(x,y)\big)\Big|_{\,y=0}
       =\left(-2\,\RB^{\za\z\zb\z}_{\z n\z n}
       -6\, k^{\za\zm}k_{\zm}^{\z\zb}\right)\nabla_{\za}\nabla_{\zb}
       + O[\CMdim{3}].
     \label{Fv_Structures}
    \end{eqnarray}
Here we everywhere omit the argument $x$ of all quantities located
on the brane, $\nabla_\alpha$ is the $d$-dimensional
covariant derivative on the brane, $k_{\alpha\beta}$ denotes the extrinsic curvature of
the brane, $k=g^{\alpha\beta}k_{\alpha\beta}$ is its trace. Subscript $n$ denotes the projection of the relevant bulk index to the normal vector $n_\ZA$, in particular $ \RB^{}_{ n n}= \RB^{\,\ZC}_{\Z \ZA\, \ZC\, \ZB} n^\ZA n^\ZB\,$.

Another source of differential operators with coefficients of
growing power in the curvatures and their derivatives is the
commutation of $\Box$ with all $x$-dependent quantities involved,
like
    \begin{eqnarray}
     && [\,\brBox,\Fu_0\,]= [\brBox,k(x)]=2\,
     (\nabla^{\za}k)\nabla_{\za} +O[\CMdim{3}]
     \nonumber\\
     && [\,\brBox,\Fv_1\,]=4\,(\nabla^\zm k^{\za\zb} )\,
     \nabla_\zm \nabla_\za \nabla_\zb + O[\CMdim{3}].
     \label{Fv_Commuted_Structures}
    \end{eqnarray}

In all these equations $O[\CMdim{l}]$ denotes the accuracy in powers
of the dimensionful quantities --- curvatures, operator potential
terms and their derivatives --- with which the relevant quantity is
calculated within the local Schwinger-DeWitt technique. In fact
$l-1$ associated with $O[\CMdim{l}]$ above indicates the
dimensionality (in mass units) of the coefficient of the
higher-derivative term of the relevant operator --- its symbol. In
\cite{PhysRep} it was called a {\em background dimensionality}, in
contrast to a total dimensionality of the quantity. In what follows
we will denote the background dimensionality of the operator
$O=O[\CMdim{l}]$ by
    \begin{eqnarray}
    {\rm Dim}\, O[\CMdim{l}]=l.
    \end{eqnarray}
Thus for an operator of the form $(\nabla^a R^m P^l k^n )\nabla^b$,
where the coefficient of the $b$-th order derivative is given by a
monomial of curvatures and potential terms and their covariant
derivatives of the total order $a$, the background dimensionality is
given by
    \begin{eqnarray}
    {\rm Dim}\,(\nabla^a R^m P^l k^n)\nabla^b =
    {\rm Dim}\,O(\CMdim{a+2m+2l+n})=a+2m+2l+n,
    \end{eqnarray}
whereas its total dimensionality is, of course, $[(\nabla^a R^m P^l
k^n)\nabla^b]=a+b+2m+2l+n$.

The background dimensionality in fact counts the order of the
perturbation theory in powers of the local quantities
$R=O(\CMdim{2})$, $P=O(\CMdim{2})$, $k=O(\CMdim{1})$ and their
derivatives. The orders of this perturbation theory have the form
$\CMdim{N}/M^M$, where $M$ is the bulk mass playing the role of the
cutoff, $\CMdim\ll M$, beyond which the local expansion does not
apply. The background dimensionalities of relevant local operators,
like (\ref{Fv_Structures})-(\ref{Fv_Commuted_Structures}), always
contribute to $N$. On the contrary, free (acting to the right)
derivatives of these operators may effect the overall power of the
cutoff $M$ in the denominator (by reducing it) so that they are not
indicative of the accuracy of the Schwinger-DeWitt expansion. In
other words, any quantity of the background dimensionality $l$
contributes to the $l$-th order of the local expansion $\CMdim{l}$
and higher, whereas the total dimensionality of this quantity does
not determine the order of this expansion. This explains a
distinguished role of the {\em background dimensionality} vs the
total one.

The next calculational step consists in the substitution of
(\ref{G0}) and (\ref{FBdev_y_expansion}) into (\ref{Gpert}), and it
leads to exactly calculable integrals over $y$. The integration over
$y$ results in a nonlocal series in inverse powers of $\rt$ --- this
is obvious from the $y$-expansion of (\ref{FBdev_y_expansion}),
because every extra power of $y$ brings one extra inverse power of
$\rt$. Each $k$-th order of this series arises in the form of the
following nonlocal chain of square root ``propagators",
    \begin{eqnarray}
    \frac1{(\rt)^{l_1}}\,V_1\frac1{(\rt)^{l_2}}\,V_2...
    V_{p{-}1}\frac1{(\rt)^{l_p}},\,\,\,\,\,
    l_1+l_2+...+l_p=k,\nonumber
    \end{eqnarray}
with some differential operators $V_i$ as its vertices. With the aid
of the commutation relations like (\ref{Fv_Commuted_Structures}) all
these propagators can be systematically commuted to the right of the
expression by the price of extra commutator terms of the same
structure, and the perturbation expansion finally takes the form
    \begin{eqnarray}
    -\big[ \overrightarrow{W} G_{D}(y,y')
    \overleftarrow{W} \big]_{\,y=y'=0}
    = \rt \;-\; \sum\limits_{k=1}^\infty
    \HU_k(\nabla)\frac1{(\rt)^{k{-}1}},    \label{temp}
    \end{eqnarray}
where $U_k(\nabla)$ is a set of certain local covariant differential
operators acting on the brane. The dimensionality of each
$\HU_k(\nabla)$ is the inverse length to the power $k$, which is
composed of the dimensionalities of bulk and extrinsic curvatures
and covariant derivatives all taken on the brane at $y=0$.

With  $\Fbr(\nabla)$ given by (\ref{hFbr_def}) the brane-to-brane
operator (\ref{5}) reads
    \begin{eqnarray}
     &&\Feff(\nabla)
     ={\Feffo}
       -\sum\limits_{k=1}^\infty
       \HU_k(\nabla)
       \frac1{(\rt)^{k{-}1}}, \label{Feff_HU_Expansion}\\
       &&{\Feffo}
      =\frac1{2\Mdgp}\Big(-\brBox+\Mbr^2+2\Mdgp \rt\Big),
     \label{Feffo_def}
    \end{eqnarray}
Here we absorbed the potential term $p$ of the operator
$\Fbr(\nabla)$ into the first term, $k=1$, of the perturbation part
of $\Feff(\nabla)$ by redefining the $\HU_1$ term,
${\HU}_1\to\HU_1-\Pbr/2\Mdgp$, because $p$ should of course be
treated on equal footing with other perturbations\footnote{We do not
introduce a new notation for $\HU_1$, and this should not lead to a
confusion because Eq.(\ref{temp}) will not be used in what
follows.}. The rest of the perturbation part is induced from the
bulk and does not depend on the boundary conditions on the brane
encoded in the operator $\Fbr(\nabla)$. Note that the zeroth-order
term here is a nontrivial nonlocal operator because $\Box$ is a
curved space $d$-dimensional D'Alembertian acting on the brane.

In the $\CMdim{2}$ approximation (involving the terms linear in the
bulk curvature and potential $P$ and the terms quadratic in the
extrinsic curvature $k$ and the brane potential $p$) the operator
coefficients $\HU_k(\nabla)$ extend to $k=6$. Higher order
coefficients go beyond the $\CMdim{2}$-approximation. The
calculations show that for a single scalar field they read as
    \begin{eqnarray}
     &&\HU_1(\nabla)
      =- \frac12\,\Fu_0 - \frac{\Pbr}{2\Mdgp}
      =-\frac12\,k - \frac{\Pbr}{2\Mdgp}
             \nonumber\\
             &&\nonumber\\
     &&\HU_2(\nabla)
     =- \frac{1}{4}\,\Fu_1 -\frac{1}{2}\,\Fv_0  -\frac{1}{8}\,\Fu_0\,\Fu_0
     =\frac{1}{4}\,\RB_{n n } + \frac{1}{4}\,k^2_{\zm\zn}
       -\frac{1}{2}\,\PB -\frac{1}{8}\,k^2,
             \nonumber\\
             &&\nonumber\\
     &&\HU_3(\nabla)
     =- \frac{1}{8}\,\Fu_2
     -\frac{1}{8} \,[\Fu_0\,, \Fv_0]
     -\frac{1}{8}\,[\,\brBox\,,\Fu_0]
     -\frac{1}{8}\,\Fu_1\, \Fu_0
     -\frac{1}{4}\,\Fv_1
     \nonumber\\
      &&\qquad\qquad\qquad\qquad\qquad= -\frac{1}{2}\,k^{\za\zb}\,\nabla_{\za}\nabla_{\zb}
          -\frac{1}{2}\,{k^{\za\zb}}_{;\za} \nabla_{\zb}
           +  O(\CMdim{3}),             \nonumber\\
     &&\HU_4(\nabla)
     =- \frac18\, \Fv_2 + O(\CMdim{3})
     =\left(\frac14\,\RB^{\za\z\zb\z}_{\z n\z n}
     +\frac34\, k^{\za\zm}k_{\zm}^{\z\zb}\right) \nabla_{\za}\nabla_{\zb}
       + O(\CMdim{3}),\nonumber\\
             &&\nonumber\\
     &&\HU_5(\nabla)     
     =- \frac18\,[\,\brBox\,,\Fv_1] + O(\CMdim{3})
     = - \frac12(\nabla^\zm k^{\za\zb} )\,\nabla_\zm \nabla_\za \nabla_\zb
     + O(\CMdim{3}),
             \nonumber\\
     &&\HU_6(\nabla)
     =\frac5{32}\,\Fv_1\, \Fv_1
            + O(\CMdim{3})=
            \frac58\, k^{\za\zb}k^{\zm\zn} \nabla_{\za}\nabla_{\zb}\nabla_{\zm}\nabla_{\zn}
       + O(\CMdim{3}).
        \label{HU_Summary}
    \end{eqnarray}
As mentioned above, each $\HU_k$ has a total dimensionality $k$ in
units of mass. Except the case of $k=2$, for which
$\HU_2=O(\CMdim{2})$, their background dimensionality is
$\HU_k=O\big(\CMdim{[(k{+}2)/3]}\big)$, where the square brackets
denote the integer part of a fractional number.

\section{Perturbation theory for the brane effective action}
\hspace{\parindent} Perturbation theory for the effective action
immediately follows from the perturbation series
(\ref{Feff_HU_Expansion}) for the operator $\Feff$. The brane
effective action can be rewritten as
    \begin{eqnarray}
     &&\frac12\;\mathrm{Tr}\ln{\Feff}
     =\frac12 {\rm Tr}\ln{\Feffo}
     +\frac12{\rm Tr}\ln\left(1-\sum\limits_{k\geq 1}
     U_k(\nabla)
     \frac1{(\rt)^{k-1}}\, \frac1{\Feffo}\right),  \label{1000}
    \end{eqnarray}
and reexpanded in powers of the perturbation series term under the
logarithm sign. After commuting all the square root ``propagators"
$1/(\rt)^k$ and the propagators $1/\Feffo$ to the right this
expansion takes the form
    \begin{eqnarray}
     &&\frac12\;\mathrm{Tr}\ln{\Feff}
     =\frac12 {\rm Tr}\ln{\Feffo}
     -\frac12\sum\limits_{k\geq 0,\,l\geq 1}\!
     \mathrm{Tr}\; \HW_{kl}(\nabla)
     \frac1{(\rt)^{k}}\,
     \frac1{\Feffo^{l}},      \label{TrLnFeff_HWklp_Expansion}
    \end{eqnarray}
with a new set of local covariant differential operators
$\HW_{kl}(\nabla)$ acting on the brane. For dimensional reasons the
total dimensionality of $\HW_{kl}(\nabla)$ is $k{+}l$ in units of
mass, because $\Feffo$ like $\rt$ has a unit dimensionality.

These operators are composed of the products of the operators
$U_k(\nabla)$ introduced above and their multiple commutators with
the the ``propagators" $1/(\rt)^k$ and $1/\Feffo$. These commutators
are based on the multiple use of the formula
    \begin{eqnarray}
     &&f(\brBox) B-B f(\brBox)
     = \sum\limits_{k=1}^{\infty}\frac{1}{k!}\;
     \ad^{k}_{_\brBox}B \cdot \partial^k_{_\brBox}
     f(\brBox),\,\,\,\,\,\ad_{_\brBox}B\equiv[\brBox,B],
    \end{eqnarray}
which leads to the following structure of $\HW_{kl}(\nabla)$
    \begin{eqnarray}
     \HW_{kl}(\nabla)=\!\!
     \sum\limits_{p=0}^{\max\{0,\,l{-}2\}}\!\!
     \frac1{\Mdgp^p}\,\HW_{kl,\;p}(\nabla).   \label{Wklp}
    \end{eqnarray}
Here, modulo the powers of $p/2\Mdgp$ originating from
$U_1(\nabla)$, the coefficients $\HW_{kl,\;p}(\nabla)$ are
$\Mdgp$-independent, and the negative powers of $\Mdgp$ follow from
the differentiation of the propagator $1/\Feffo$ with respect to
$\Box$ participating in the commutators $[U_k,1/\Feffo]$,
    \begin{equation}
     \partial_{_\brBox}\,\frac{1}{\Feffo}\;=\; \frac{1}2\,
     \frac{1}{\Feffo^{2}}
     \left(\frac1{\Mdgp}+\frac1{\rt}\right).         \nonumber
    \end{equation}
As we see, each such differentiation results in the extra power of
the propagator $1/\Feffo$ and an extra term proportional to
$1/\Mdgp$. For a $\HW_{kl}(\nabla)$ with a given $l$ the highest
order of $1/\Mdgp$ is limited by $\max\{0,\,l{-}2\}$. This is
because $p$ differentiations increase the power of $1/\Feffo$ from
some initial $l'$ to at least $l=l'+p$, and the initial $l'\geq 2$,
because the commutation of $1/\Feffo$ with other quantities begins
only when the number of these propagators exceeds two (one should
remember that one propagator $1/\Feffo$ always stands to the right
of everything else, see Eq.(\ref{1000})). This explains the upper
limit of summation over $p$ in (\ref{Wklp}) and, as we will later
see, underlies the regularity of the Neumann limit $\Mdgp\to
\infty$.

The background dimensionality of $\HW_{kl,\;p}(\nabla)$ is a
monotonically growing function of all its indices and can be shown
to satisfy the bound
    \begin{eqnarray}
    {\rm Dim}\, \HW_{kl,\;p}(\nabla)\geq
    \left[\frac{k+l+2p+2}{3}\right],      \label{bound}
    \end{eqnarray}
where square brackets denote an integer part of the fractional
number. This bound follows from the observation that for any
$\HW_{kl,\,p}$, composed of the chain of $U_{k_i}$ and $n$
commutators with $\Box$, the total dimensionality
$[\HW_{kl,\,p}\,]=k{+}l{+}p$ equals $[\HW_{kl,\,p}\,]=\sum_i
k_i+2n$. On the other hand, the background dimensionality in
addition to the sum of ${\rm Dim}\,U_{k_i}=[(k_i+2)/3]$ contains at
least one extra unit of mass per each commutation with $\Box$.
Therefore
    \begin{eqnarray}
    {\rm Dim}\,\HW_{kl,\,p}(\nabla)\geq
    \sum_i \left[\frac{k_i{+}2}3\right]+n  \geq
    \left[\frac{\sum_i k_i+2}3\right]+n=
    \left[\frac{(k{+}l{+}p){+}n{+}2}3\right],
    \end{eqnarray}
where we used the above counting of the total dimensionality $\sum_i
k_i+2n=k{+}l{+}p$. The bound (\ref{bound}) then follows from the
fact that the overall negative power of $\Mdgp$ does not exceed the
number of commutations $n$, $0\leq p\leq n$.

Relevant $\HW_{kl}$ which can contribute up to
$O\big(\CMdim{2}\big)$ order inclusive are
    \begin{eqnarray}
     &&\HW_{01}=\HU_1
     = - \fract12 k-\fract{\Pbr}{2\Mdgp},
\nonumber\\
\nonumber\\
     &&\HW_{11}=\HU_2
     =\fract{1}{4}\,\RB_{nn} + \fract{1}{4}\,k^2_{\zm\zn}
       -\fract{1}{2}\,\PB -\fract{1}{8}\,k^2,
     \nonumber
        \end{eqnarray}
    \begin{eqnarray}
     &&\HW_{02}=\fract12\,\HU^2_1
      =
      \fract12\,\left(\fract12 k + \fract{\Pbr}{2\Mdgp} \right)^2,
\nonumber\\
\nonumber\\
     &&\HW_{21}=\HU_3
      =- \fract{1}{2}\,k^{\za\zb}\,\nabla_{\za}\nabla_{\zb}
           - \fract{1}{2}\,(\nabla_\alpha k^{\za\zb}) \nabla_{\zb}
           +  O(\CMdim{3}),
\nonumber\\
\nonumber\\
     &&\HW_{31}=\HU_4
      =\left(\fract14\,\RB^{\za\z\zb\z}_{\z n\z n}
      +\fract34\, k^{\za\zm}k_{\zm}^{\z\zb}\right)\nabla_{\za}\nabla_{\zb}
       + O(\CMdim{3}),
\nonumber\\
\nonumber\\
     &&\HW_{22}=\HU_3{\HU}_1+\fract12\HU_2\HU_2
      =
      \fract12\, \left(\fract12 k + \fract{\Pbr}{2\Mdgp} \right) \,k^{\za\zb}\,\nabla_{\za}\nabla_{\zb}
       + O(\CMdim{3}),
\nonumber\\
\nonumber\\
     &&\HW_{41}=\HU_5
      =- \fract12( \nabla^\zm k^{\za\zb} )\,\nabla_\zm \nabla_\za \nabla_\zb
            + O(\CMdim{3}),
\nonumber\\
\nonumber\\
     &&\HW_{51}=\HU_6
      =\fract58\, k^{\za\zb}k^{\zm\zn}
      \nabla_{\za}\nabla_{\zb}\nabla_{\zm}\nabla_{\zn}
       +O(\CMdim{3}),
\nonumber\\
\nonumber\\
     &&\HW_{42}=\HU_5 {\HU}_1 + \HU_4 \HU_2
     + \fract12 \HU_3 \HU_3 + \HU_3 [\,\brBox,{\HU}_1]
      =
      \fract18\, k^{\za\zb}k^{\zm\zn}
      \nabla_{\za}\nabla_{\zb}\nabla_{\zm}\nabla_{\zn}
       + O(\CMdim{3}).
     \label{HW_Summary}
    \end{eqnarray}

 \section{Generalized proper time method.}

 The further calculation is based on the possibility to express the
 nonlocal structures in (\ref{TrLnFeff_HWklp_Expansion}) in terms of
 the heat kernel of the box operator $\Box$, which admits a
 well-known curvature expansion. This is the set of proper time
 representations which differ from a usual Schwinger integral by
 nontrivial weight functions $w_{kl}(s)$
  \hspace{\parindent}
     \begin{eqnarray}
      && \ln\Feffo
      \;=\;{-}\ln(2\Mdgp)-\int\limits_0^\infty \frac{d\pt}{\pt} \,
      \MW_{00}(\pt) \,
      e^{-\pt (\MB^2-\brBox)}\;,                   \label{MW_00_def}\\
      && \frac1{(\rt)^{k}}\, \frac1{\Feffo^{l}}
      \;= \int\limits_0^\infty \frac{d\pt}{\pt}\;
      \MW_{kl}(\pt)\,e^{-\pt(\MB^2-\brBox)}\,,
      \qquad\qquad (l\geq1)\;.                       \label{MW_kl_def}
     \end{eqnarray}

These weight functions can be found as follows. First, decompose the
operator $\Feffo$ defined by (\ref{Feffo_def}) into the product of
two factors linear in $\rt$
    \begin{eqnarray}
     &&\Feffo
      =\frac1{2\Mdgp}(\rt-\Mp)(\rt-\Mm)  \;,
     \label{Feffo_Decomposition}
    \end{eqnarray}
where $\Mx_\pm$ denote the roots of the quadratic equation
$x^2+2\Mdgp x-\MB^2+\Mbr^2=0$,
    \begin{eqnarray}
     \Mp=-\Mdgp+\sqrt{\Mdgp^2+\MB^2-\Mbr^2},
     \qquad \Mm=-\Mdgp
     -\sqrt{\Mdgp^2+\MB^2-\Mbr^2}.     \label{roots_mpm}
    \end{eqnarray}
Therefore
     \begin{eqnarray}
      &&\ln{\Feffo}=-\ln{2\Mdgp} + \ln{(\rt-\Mp)} +
      \ln{(\rt-\Mm)},\nonumber\\
      &&\nonumber\\
      &&
     \fracd{1}{(\rt)^k\Feffo^{l}}
       =\sum\limits_{a=1}^{k}
       \frac{B^a_{kl}(\Mp,\Mm)}{(\rt)^{a}}\nonumber\\
         &&\qquad\qquad\qquad\qquad\qquad
         + \sum\limits_{b=1}^{l}\left(
         \frac{D^b_{kl}(\Mp,\Mm)}{(\rt\!-\!{\Mp})^{\,b}}
         +(m_+\leftrightarrow
         m_-)\right).   \label{MW_kl_PartialFractions_decomposition}
     \end{eqnarray}
Here the second equation is the result of the decomposition of its
left hand side into partial fractions with the coefficients
$B^a_{kl}$ and $D^b_{kl}$
    \begin{eqnarray}
     && B^a_{kl}{(\Mp,\Mm)}=
     \frac{(-\Mdgp/2)^{l}}{(\hdM)^{2l}}
                             \frac{1}{(l{-}1)!\,(k{-}a)!}\nonumber\\
     &&\qquad\qquad
     \times\sum\limits_{b=1}^{l}
                             \frac{\GammaF{2l{-}b}\GammaF{k{-}a{+}b}}
                             {\GammaF{l{-}b{+}1}\GammaF{b}}
                             \left( \frac{1}{\Mp^{k-a}}\;
                             \left(\frac{\Mp\!{-}\Mm}{\Mp}\right)^b
                             + \fract{1}{\Mm^{k-a}}\;
                             \left(\fract{\Mm\!{-}\Mp}{\Mm}\right)^b
                             \right),
    \label{PartialFractions_BMo_coef}
    \end{eqnarray}
and
    \begin{eqnarray}
     D^b_{kl}{(\Mp,\Mm)}= \left(\frac{\Mdgp}{\hdM}\right)^{l} \;
     \frac{1}{\GammaF{k}\GammaF{l}}
                             \; \frac{(-1)^{l-b}}{\Mp^{k+l-b}}
                             \;\;\sum\limits_{p=0}^{l-b}
                             \frac{\GammaF{k{+}l{-}b{-}p}}{\GammaF{l{-}b{+}1{-}p}}
                             \frac{\GammaF{l{+}p}}{p!}
                             \left(\fract{\Mp}{\Mp\!{-}\Mm}\right)^{p}.
    \label{PartialFractions_BMpm_kl_coefs}
    \end{eqnarray}

Now, in addition to a simple proper time representation of the
square root ``propagators"
     \begin{eqnarray}
      &\fracd{1}{(\rt)^a}
       &=\frac{1}{\GammaF{ a/2}}
         \int\limits_0^\infty d\pt
         \;\pt^{a/2{-}1} \,
         e^{-\pt (\MB^2-\brBox)} \;,    \label{Pure_Neumann}
    \end{eqnarray}
we need a similar representation for the new operator of the form
$\rt {-} \Mx$ and its logarithm. It can be derived with the aid of
the integral representation of the cylindrical function of a
half-integer order
         \begin{eqnarray}
      &\fracd{1}{(\rt {-} \Mx)^\alpha}
       &=\frac{1}{\GammaF{\alpha}}\int\limits_0^\infty {d x} \;
       x^{\alpha{-}1} \,e^{\Mx x}\;e^{-x\rt}
       \nonumber\\
       &&=\frac{\alpha}{2\sqrt\pi \GammaF{1+\alpha}}
        \int\limits_0^\infty {d\pt} \;\pt^{-3/2} \;  e^{-\pt (\MB^2-\brBox)}
        \int\limits_0^\infty {d x} \;x^\alpha \;e^{\Mx x}\;
        \;e^{-\frac{x^2}{4\pt}}.
    \end{eqnarray}
Differentiating it with respect to $\alpha$ at $\alpha=0$
one gets
         \begin{eqnarray}
      &\ln(\rt {-} \Mx)
       &=-\frac1{2\sqrt\pi}
        \int\limits_0^\infty {d\pt} \;\pt^{-3/2} \;  e^{-\pt (\MB^2-\brBox)}
        \int\limits_0^\infty {d x} \;e^{\Mx x}\;
        \;e^{-\frac{x^2}{4\pt}}\nonumber\\
        &&=
        {-}\frac12\int\limits_0^\infty
        \frac{d\pt}{\pt} \,
      w(-m\sqrt\pt) \,e^{-\pt (\MB^2-\brBox)}, \label{1003}
    \end{eqnarray}
where the function $w(-m\sqrt\pt)$ is given in terms of the
complementary error function $\erfc(z)$,
    \begin{eqnarray}
     && w(-\sigma)\equiv
     \frac{2}{\sqrt\pi}\;\int\limits_0^\infty {d {x}} \;e^{2\sigma{x}}\;e^{-{x}^2}
      = e^{\sigma^2} \erfc{({-}\sigma)},\\
     &&\erfc(z)=\frac2{\sqrt{\pi}}\int\limits_z^\infty dt\;e^{-t^2}.
    \end{eqnarray}
A multiple differentiation of (\ref{1003}) with respect to $m$ then
gives
     \begin{eqnarray}
      &\fracd{1}{(\rt {-} \Mx)^a}
       &=\frac{1}{2\GammaF{a}}
         \int\limits_0^\infty d\pt \;\pt^{a/2{-}1}  \;
       \left.\frac{d^a w(-\sigma)}{d\sigma^a}\,
       \right|_{\,\sigma=\Mx\sqrt{\pt}}
       \;e^{-\pt (\MB^2-\brBox)}.    
      \label{Pure_Robin_sum_erfc}
    \end{eqnarray}
Using (\ref{Pure_Neumann}), (\ref{1003}) and
(\ref{Pure_Robin_sum_erfc}) in
(\ref{MW_kl_PartialFractions_decomposition}) we finally come to the
following expressions for the weights in
Eqs.(\ref{MW_00_def})-(\ref{MW_kl_def})
    \begin{eqnarray}
    &&\MW_{00}= \frac1{2}\,\Big(\,
      w(-\Mp\sqrt{\pt})+w(-\Mm\sqrt{\pt})\,\Big),   \label{MW_00_res}\\
     &&\MW_{kl}
      =\sum\limits_{a=1}^{k}
      \frac{{B^a_{kl}(\Mp,\Mm)}}{2\GammaF{a/2}}\;\pt^{a/2}
      \nonumber\\
      &&\qquad
         + \sum\limits_{a=1}^{l}
         \frac{{D^a_{kl}(\Mp,\Mm)}}{2\GammaF{a}}\;
         \pt^{a/2}
         \left.\frac{d^a w(-\sigma)}{d\sigma^a}
         \right|_{\,\sigma=\Mp\sqrt{\pt}}
         \;+\;\{\Mp\leftrightarrow\Mm\}\;.  \label{MW_kl_SumErfc_Rep}
    \end{eqnarray}

\newcommand{\sm}{{-}} 
In terms of these weights the action
(\ref{TrLnFeff_HWklp_Expansion}) takes the form
    \begin{eqnarray}
     &\frac12\;\mathrm{Tr}\ln{\Feff}
     &\equiv - \frac12\sum\limits_{k,\,l=0}^\infty\;
     \int\limits_0^\infty \frac{d\pt}{\pt}\;\MW_{kl}(\pt)\,
     e^{-\pt\MB^2}\,
      \mathrm{Tr} \left(\HW_{kl}(\nabla)
      \;e^{\pt\brBox}\right),             \label{TrLnFeff_HW_PT}
    \end{eqnarray}
where we disregarded the contribution of the local measure $-{\rm
Tr}\,\ln(2\Mdgp)\sim \delta^{(d)}(0)$ and, in addition to $\HW_{kl}$
of (\ref{TrLnFeff_HWklp_Expansion}), introduced
    \begin{eqnarray}
    \HW_{00}=1,\,\,\,\,\HW_{k0}=0,\,\,k{\geq}1. \label{Wk0}
    \end{eqnarray}

\section{Generalized Schwinger-DeWitt expansion.}
 \hspace{\parindent}
The calculation of (\ref{TrLnFeff_HW_PT}) is based on the heat
kernel expansion (\ref{braneHK_Ansatz}) for the covariant
d'Alember\-tian $\Box$ acting in a curved $d$-dimensional space
without boundaries. The Schwinger-DeWitt coefficients $a_n(x,x')$ in
this expansion represent brane curvature invariants of the growing
background dimensionality $2n$, $a_n(x,x')=O(\CMdim{2n})$. With this
expansion the functional traces in (\ref{TrLnFeff_HW_PT}) take the
form
    \begin{eqnarray}
     \mathrm{Tr}\Big(\HW_{kl}(\nabla)\,e^{\,\pt\,\brBox}\Big)
     = \frac1{(4\pi \pt)^{d/2}}
       \sum_{n=0}^\infty\,\pt^n\,
       \brint \;\HW_{kl}\big(\tilde{\nabla}(\pt)\big)\,
       a_n(x,x')\,\Big|_{\,x'=x}\;,
     \nonumber
    \end{eqnarray}
where the new generalized covariant derivative
    \[\tilde{\nabla}_\za(\pt)
      \equiv \nabla_\za -\fract1{2\pt}\partial^x_\za\brWF(x,x')
    +\fract12\partial^x_\za\ln D(x,x')\]
originates from the commutation of $\nabla_\alpha$ with the
exponential factor $\exp(-\brWF(x,x')/2\pt)$ and the Van
Vleck-Morette determinant $D(x,x')$ in the kernel of
$\exp(\pt\brBox)$. In addition to
\[\nabla_{\zm_1}...\nabla_{\zm_p}\,a_{n}(x,x')\Big|_{x'=x}\]
this lengthening of the covariant derivatives also generates the
coincidence limits
\[\nabla_{\zm_1}...\nabla_{\zm_p}\brWF(x,x')\Big|_{x'=x},\,\,\,\,
\nabla_{\zm_1}...\nabla_{\zm_p}D(x,x')\Big|_{x'=x}\] easily
calculable by the DeWitt recurrence procedure \cite{DeWitt,PhysRep}.
Moreover, this leads to extra negative powers of the proper time, so
that
    \begin{eqnarray}
     \brint \HW_{kl}\big(\tilde{\nabla}^{x}\big)
     \;a_{n}(x,x')\,\Big|_{x=x'}
       =
     \sum\limits_{p=0}^{\max\{0,\,l{-}2\}}\,
     \sum\limits_{c=0}^{\left[\fracs{2k+2l+p}{6}\right]}\,
     \frac{(\Abr_n)^c_{kl,p}}{\pt^c\;\Mdgp^p} \;,   \label{Abr_bc_kl_def}
    \end{eqnarray}
where $(\Abr_n)^c_{kl,p}$ represents the set of integrals of bulk
and brane curvature invariants, powers of potential terms $\PB$ and
$\Pbr/m$ and their derivatives of the growing dimensionality
$k{+}l{+}p+2n{-}2c$ (which now coincides with their background
dimensionality, because they are no longer the differential
operators),
    \begin{eqnarray}
     (\Abr_n)^c_{kl,\,p}=O\big(\CMdim{k+l+p+2n-2c}\big).
    \end{eqnarray}
The highest power of $1/s$ in (\ref{Abr_bc_kl_def}) is determined by
one half of the order of $\HW_{kl,p}(\nabla)$ in derivatives and
equals the difference between the total dimensionality of this
operator and its background dimensionality. As $\HW_{kl}$ is a sum
over the powers of the inverse DGP mass scale (\ref{Wklp}), the
quantities $(\Abr_n)^c_{kl,\,p}$ above represent the relevant
coefficients of $1/\Mdgp^p$.

Thus
    \begin{eqnarray}
     &&\!\!\!\!\frac12\;\mathrm{Tr}\ln{\Feff}
     =- \frac12\frac{1}{(4\pi)^{d/2}}\!\!
     \sum\limits_{\,\{k,\,l,\,n,\,p,\,c\}\,}
       \int\limits_0^\infty \frac{d\pt}\pt\;\pt^{n-c-d/2}\,
     \MW_{kl}(\pt)\;
     e^{-\pt\MB^2}\,
     \frac{(\Abr_n)^{c}_{kl,\,p}}
     {\Mdgp^p} \;,           \label{TrLnFeff_GenSDW_Expansion}
    \end{eqnarray}
where the domain of summation over all indices is given by
    \begin{eqnarray}
     \sum\limits_{\,\{k,\,l,\,n,\,p,\,c\}\,}=
     \sum\limits_{k,\,l,\,n=0}^\infty
     \sum\limits_{p=0}^{\max\{0,\,l{-}2\}}\,
     \sum\limits_{c=0}^{\left[\fracs{2k+2l+p}{6}\right]}
    \end{eqnarray}

The quantities $(\Abr_n)^{c}_{kl,\,p}$ play the role of integrated
generalized Schwinger-DeWitt coefficients. For a scalar field in
the approximation $O\big(\CMdim{2}\big)$ only the following $p=0$
coefficients contribute to the brane effective action
    \begin{eqnarray}
     & (\Abr_0)^{0}_{00,0}
      &= \brint,
    \nonumber\\
     & (\Abr_1)^{0}_{00,0}
      &= \brint \;\,\fract16\,\Rbr(g)\;\;
       = \brint \;
         \left(\fract16\RB - \fract13\RB_{nn}
         + \fract16\,k^2 - \fract16\,k^2_{\za\zb}
         \right),
    \nonumber\\
     & (\Abr_0)^{0}_{01,0}
      &= \brint \;\left(\,{-}\fract12 k - \fract{\Pbr}{2\Mdgp} \right),
    \nonumber\\
     & (\Abr_0)^{0}_{11,0}
      &= \brint \; \left(\;\fract{1}{4}\RB_{nn}+ \fract{1}{4}\,k^2_{\zm\zn}
                        -\fract{1}{2}\,\PB-\fract{1}{8}\,k^2\right),
    \nonumber\\
     & (\Abr_0)^{0}_{02,0}
      &= \brint \; \left(\;\fract18 \,k^2 + \fract12 \,k\,\fract{\Pbr}{2\Mdgp}
                        +\fract12\Big(\fract{\Pbr}{2\Mdgp}\Big)^2\,\right),
      \nonumber\\
    & (\Abr_0)^{1}_{21,0}
     & =  \brint \; \, \fract{1}{4}\,k,
    \nonumber\\
    & (\Abr_0)^{1}_{31,0}
     & =  \brint \; \left(\,{-}\fract18\,\RB_{nn} +\fract38\,
     k^2_{\za\zb}\right),
    \nonumber\\
    & (\Abr_0)^{1}_{22,0}
     & =  \brint \; \left({-}\fract18 k^2 -\fract14
     \,k\,\frac{\Pbr}{2\Mdgp}\right),
    \nonumber\\
    & (\Abr_0)^{2}_{51,0}
     & =  \brint \; \left(\fract5{16}\,
     k^2_{\za\zb}+\fract5{32}\,k^2\right).      \label{Abr_Summary}
    \end{eqnarray}

\section{Large mass expansion and its cutoff scales}
 \hspace{\parindent}
Integration over $\pt$ in (\ref{TrLnFeff_GenSDW_Expansion}) gives
    \begin{eqnarray}
     &\frac12\;\mathrm{Tr}\ln{\Feff}
      &=- \frac12\frac{M^d}{(4\pi)^{d/2}}
      \sum\limits_{\,\{k,\,l,\,n,\,p,\,c\}\,}
     \MC^{n{-}c}_{kl}
     \frac{(\Abr_n)^{c}_{kl,\,p}}
     {M^{2n-2c+k+l}\;\Mdgp^p},      \label{TrLnFeff_sum_Cklnu_Aklnu}
    \end{eqnarray}
where $\MC^j_{kl}$ (with $j=n-c$) are the following functions of the
mass parameters of the model
    \begin{eqnarray}
     &&\MC^j_{kl}
     = M^{2j-d+k+l}
     \int\limits_{0}^{\infty}\frac{d\pt}{\pt}\;
     \pt^{j-d/2}\,\MW_{kl}(\pt)\;e^{-\pt\MB^2}\;.   \label{MC_klnu_def}
    \end{eqnarray}

The behavior of these functions for $\MB\to\infty$ is important for
the determination of the efficiency of the expansion
(\ref{TrLnFeff_sum_Cklnu_Aklnu}) and of the range of its validity
--- the cutoff $\MB_{\rm cutoff}$ below which, $\CMdim\ll\MB_{\rm
cutoff}$, this expansion makes sense. This behavior easily follows
from a simple observation that the functions $\MC^j_{kl}$ can be
directly obtained from the nonlocal form-factors (\ref{MW_kl_def})
by integration over their argument $\Box=-\lambda$ with the weight
$\lambda^{d/2-j-1}$. In the domain of convergence of this integral
in the complex plane of $d$ we have
     \begin{eqnarray}
      \int\limits_0^\infty
      \frac{d\lambda}\lambda\,\lambda^{d/2-j}
      \left.\frac1{(\rt)^{k}}\,
      \frac1{\Feffo^{l}}\right|_{\,\Box=-\lambda}
      =\frac{\GammaF{d/2-j}}{M^{2j-d+k+l}}\MC^j_{kl}\;.
     \end{eqnarray}
With the replacement of the integration variable
$x=\sqrt{1+\lambda/\MB^2}-1$ the integral representation for
$\MC^j_{kl}$ takes the form
     \begin{eqnarray}
     &&\MC^j_{kl}=\frac1{\GammaF{\nu}}
     \left(\frac{2\Mdgp}M\right)^l\int\limits_0^\infty
     dx\,x^{\nu-1}(x+\varepsilon_+)^{-l}\,
     \varphi(x)\,\Big|_{\;\nu=d/2-j}\;,\nonumber\\
     &&\varphi(x)=\frac{(x+2)^{\nu-1}}
     {(x+1)^{k-1}(x+\varepsilon_-)^l}\,,    \label{Cintegral}
     \end{eqnarray}
where $\varepsilon_\pm=1-m_\pm/\MB$. For $\MB\to\infty$ the
parameter $\varepsilon_+\to 0$ ($\varepsilon_-\to 1$), and the
integral here has a nonanalytic in $\varepsilon_+$ part because of
the singularity of its integrand at $x=0$ \cite{toyDGP},
     \begin{eqnarray}
     \int\limits_0^\infty
     dx\,x^{\nu-1}(x+\varepsilon)^{-l}\,
     \varphi(x)=\varepsilon^{\nu-l}
     \frac{\varGamma(\nu)\varGamma(l-\nu)}{\varGamma(l)}
     \varphi(0)+O(1)\,.
     \end{eqnarray}
Since $\varepsilon_+\to \Mdgp/\MB$ in this limit, we have
    \begin{eqnarray}
     &&\MC^j_{kl}=
     C_1\left(\frac{\Mdgp}{\MB}\right)^{d/2-j}
     +C_2\left(\frac{\Mdgp}{\MB}\right)^l,
     \,\,\,\,\MB\to\infty.                        \label{Casymp}
    \end{eqnarray}

Thus, in view of the background dimensionality of
$(\Abr_n)^{c}_{kl,\,p}=O(\CMdim{k+l+p+2n-2c})$ the local expansion
(\ref{TrLnFeff_sum_Cklnu_Aklnu}) for the effective action takes the
form
    \begin{eqnarray}
     &&\frac12\;\mathrm{Tr}\ln{\Feff}
      =\left(\frac{M m}{4\pi}\right)^{d/2}\!\!\!
      \sum\limits_{\,\{k,\,l,\,n,\,p,\,c\}\,}
     \frac{\Mdgp^{c-n-p}\;
     O(\CMdim{k+l+p+2n-2c})}{M^{n-c+k+l}\;}\nonumber\\
     &&\qquad\qquad\qquad\quad
     +\frac{M^d}{(4\pi)^{d/2}}\!\!\!
      \sum\limits_{\,\{k,\,l,\,n,\,p,\,c\}\,}
     \frac{\Mdgp^{l-p}\;
     O(\CMdim{k+l+p+2n-2c})}
     {M^{2n-2c+k+2l}\;}\,.    \label{TrLnFeff_sum_Cklnu_Aklnu1}
    \end{eqnarray}
By introducing the summation index $N=n-c+k+l$ --- an overall power
of $1/\MB$ --- in the first sum and correspondingly the summation
index $L=2n-2c+k+2l$ in the second sum, one can rewrite this series
in the form (\ref{efficiency0}) presented in Introduction
    \begin{eqnarray}
     &&\frac12\;\mathrm{Tr}\ln{\Feff}
      =\left(\frac{M
      m}{4\pi}\right)^{d/2}\sum\limits_{N=0}^\infty
      \frac1{M^N}\;\sum\limits_{\,i\leq N\,}
     \frac{O(\CMdim{2N-i})}{\Mdgp^{N-i}}\nonumber\\
     &&\qquad\qquad\qquad\quad
     +\frac{M^d}{(4\pi)^{d/2}}
      \sum\limits_{L=0}^\infty
      \frac1{M^L}\sum\limits_{i\leq L}
     \Mdgp^i\;
     O(\CMdim{L-i})\, ,      \label{efficiency}
    \end{eqnarray}
where the coefficient of any power of $\MB$ turns out to be a {\em
finite} sum of terms of a limited order in $\CMdim{}$  --- the
background dimensionality of relevant field invariants.

The finiteness of such sums over $i=k+l-p$ in the first series of
(\ref{efficiency}) follows from the following simple argumentation.
The range of summation over $c\leq [\,(2k+2l+p)/6\,]$ in
(\ref{TrLnFeff_sum_Cklnu_Aklnu1}) is less than $[\,(2k+3l)/6\,]$
because $p<{\rm max}\{0,\,l-2\}$. Therefore,
    \begin{eqnarray}
    k+l=N+c-n\leq N+c<N+
    \left[\frac{k}3+\frac{l}2-\frac13\right]
    <N+\frac12(k+l),                        \nonumber
    \end{eqnarray}
so that $k+l<N$, and the ranges of summation over $c$, $p$ and
$i\equiv k+l-p$ indeed turn out to be limited at least by $N$.

Similarly, the finiteness of sums over $i=l-p$ in the second series
of (\ref{efficiency}) is based on the following chain of
inequalities
    \begin{eqnarray}
    k+2l=L+2c-2n\leq L+2c<L+\left[\frac{2k}3+l-\frac23\right],
    \end{eqnarray}
so that $[k/3+l]<L$, and the ranges of summation over $c$, $p$ and
$i\equiv l-p$ are again restricted from above for any given $L$.

This property is very important for the efficiency of the
perturbation theory with the cutoff scale $\MB$, because otherwise
any given order in $1/\MB$ would require an infinite series in
$\CMdim/\Mdgp$ -- the price one could have paid for the presence of
the second scale $\Mdgp$. Fortunately, for any $N$ only a finite
order $O(\CMdim{2N})$ of perturbation theory is required. This
follows from a special asymptotic behavior of the coefficient
(\ref{Casymp}) which brings extra powers of $1/\MB$ to
(\ref{TrLnFeff_sum_Cklnu_Aklnu}). The form of the asymptotics
(\ref{Casymp}) is responsible for the two series in the expansion
(\ref{efficiency}), having qualitatively different analytic behavior
in $\MB$ and $\Mdgp$, --- the property recently discovered for the
effective potential of the toy DGP model \cite{toyDGP}. Whereas the
second part is analytic in a small DGP scale $\Mdgp\to 0$, the first
``nonanalytic"  part is formally singular in this limit, and this
leads to the redefinition of the cutoff $M_{\rm cutoff}$ of the
theory, below which $\CMdim{}\ll M_{\rm cutoff}$ the local expansion
remains valid.

Indeed, despite the efficiency of the obtained expansion, in the
first series of (\ref{efficiency}) it contains negative powers of
the DGP scale $\Mdgp$ and blows up for small $\Mdgp\to 0$. This is a
typical situation of the presence of a strong-coupling scale
\cite{scale}. In fact, for $\Mdgp<\MB$ the actual cutoff is lower
than $\MB$ and is given by the expression (\ref{cutoff}), $M_{\rm
cutoff}=\sqrt{\MB\Mdgp}$, presented in Introduction (the condition
of smallness of the strongest $i=0$ term in the first series of
(\ref{efficiency}), $\CMdim{2}/\MB\Mdgp\ll 1$).

The actual calculation of the functions (\ref{MC_klnu_def}) can be
done by using the proper time weights (\ref{MW_00_res}) and
(\ref{MW_kl_SumErfc_Rep}). Since these weights imply explicit
symmetrization with respect to $m_\pm$, they take the form
    \begin{eqnarray}
     C^j_{kl}=\tilde
     C^j_{kl}(M,m_+,m_-)+
     (\,\Mp\leftrightarrow\,\Mm\,)\,.  \label{MC_Summary_yRepr_Reduced0}
    \end{eqnarray}
In the $O(\CMdim{2})$-approximation the relevant $\tilde\MC^j_{kl}$
turn out to be
    \begin{eqnarray}
     &&\tilde\MC^j_{00}=
           \ResHGFp{(0)}{2j,\,1},
     \nonumber\\
     \nonumber\\
     &&\tilde\MC^j_{01}
      = \left(\fract{\Mdgp}{\hdM}\right)
           \ResHGFp{(0)}{2j+1,\,2},
     \nonumber\\
     \nonumber\\
     &&\tilde\MC^j_{11}
      = \left(\fract{\Mdgp}{\hdM}\right)
           \ResHGFp{(0)}{2j+2,\,1},
     \nonumber\\
     \nonumber\\
     &&\tilde\MC^j_{02}
      =  \left(\fract{\Mdgp}{\hdM}\right)^{\!2}
      \left(
       {-} \fract{2\MB}{\!\hdM}
           \ResHGFp{(0)}{2j+1,\,2}
       +
           \ResHGFp{(0)}{2j+2,\,3}
      \right),
     \nonumber\\
     \nonumber\\
     &&\tilde\MC^j_{21}
       = 2\left(\fract{\Mdgp}{\hdM}\right)
           \ResHGFp{(1)}{2j+2,\,1},
     \nonumber\\
     \nonumber\\
     &&\tilde\MC^j_{31}
       =  2\left(\fract{\Mdgp}{\hdM}\right)
             \ResHGFp{(2)}{2j+2,\,1},
       \nonumber\\
       \nonumber\\
     &&\tilde\MC^j_{22}
      =  2\left(\fract{\Mdgp}{\hdM}\right)^{\!2}
          \left(
            {-}\fract{2\MB}{\hdM}
                \ResHGFp{(1)}{2j+2,\,1}
            -   \ResHGFp{(2)}{2j+2,\,1}\right.\nonumber\\
            &&\qquad\qquad\qquad\qquad\left.
            +  \ResHGFp{(1)}{2j+3,\,2}
            \vphantom{\fract{2\MB}{\hdM}}
          \right),
       \nonumber\\
     &&\tilde\MC^j_{51}
      =2\left(\fract{\Mdgp}{\hdM}\right)
               \ResHGFp{(4)}{2j+2,\,1},
       \nonumber\\
       \nonumber\\
     &&\tilde\MC^j_{42}
       =  2\left(\fract{\Mdgp}{\hdM}\right)^2
          \left(
            {-}\fract{2\MB}{\hdM}
                \ResHGFp{(3)}{2j+2,\,1}
            - 3 \ResHGFp{(4)}{2j+2,\,1}\right.
            \nonumber\\
       &&\qquad\qquad\qquad\qquad\left.
       +   \ResHGFp{(3)}{2j+3,\,2}
            \vphantom{\fract{2\MB}{\hdM}}
          \right).                   \label{MC_Summary_yRepr_Reduced}
    \end{eqnarray}
Here
    \begin{eqnarray}
    \hdM\equiv\frac{\Mp{-}\Mm}2,
    \end{eqnarray}
the basic function $\ResHGFp{(0)}{a,\,b}$ is given by the
regularized Gauss hypergeometric function
    \begin{eqnarray}
     &&\ResHGFp{(0)}{a,\,b}=
     \ResHGFp{(0)}{a,\,b\,|\,\sigma}\,
     \Big|_{\,\sigma=\Mp/2\MB}\nonumber\\
      &&\qquad\qquad\quad
      \left.\equiv \frac{\GammaF{a-d}\GammaF{b}}
    {\GammaF{\frac{a+b+1-d}2}}
    \HGF{a-d,\,b}{\fract{a+b+1-d}2}{\sigma+\frac12}
    \,\right|_{\,\sigma=\Mp/2\MB}    \label{5000}
    \end{eqnarray}
and
    \begin{eqnarray}
     &&\ResHGFp{(n)}{a,\,b}
      =\frac1{\sigma^n}
             \left.\Big(\ResHGFp{(0)}{a,\,b\,|\,\sigma}
      -\sum\limits_{k=0}^{n{-}1}
      \frac{1}{k!}\,
      \Big[\,d^k\!\ResHGFp{(0)}{a,\,b\,|\,\sigma}/d\sigma^k\,
                   \Big]_{\sigma=0}\,
            \sigma^k
             \Big)\,\right|_{\sigma=\Mp/2\MB}       \label{ResHGF_def}
    \end{eqnarray}
is the function $\ResHGFp{(0)}{a,\,b\,|\,\sigma}/\sigma^n$ with the
singular at $\sigma=0$ part subtracted, also taken at
$\sigma=\Mp/2\MB$.

The transformation property of the hypergeometric function from the
argument $z$ to $1-z$ allows one to rewrite (\ref{5000}) in the form
which reveals the structure of the expansion (\ref{efficiency})
    \begin{eqnarray}
     &&\ResHGFp{(0)}{a,\,b}=
    \left(\fract{\varepsilon_+}2\right)^{\frac{d+1-a-b}2}
      \GammaF{\fracd{a+b-1-d}2}\nonumber\\
      &&\qquad\qquad\qquad\qquad\qquad\times
    \HGF{\frac{d+1-a-b}2,\,\fract{a+b+1-d}2}
    {\fract{d+3-a-b}2}{\fract{\varepsilon_+}2}\nonumber\\
     &&\qquad\qquad\qquad
      +\frac{\GammaF{a-d}\GammaF{b}\GammaF{\frac{d-a-b+1}2}}
      {\GammaF{\frac{d+b-a+1}2}\GammaF{\frac{a-b+1-d}2}}\nonumber\\
    &&\qquad\qquad\qquad\qquad\qquad\times
    \HGF{\frac{d+1-a-b}2,\,\fract{a+b+1-d}2}
    {\fract{d+3-a-b}2}{\fract{\varepsilon_+}2},    \label{5001}
    \end{eqnarray}
where $\varepsilon_+=1-m_+/\MB$ as in (\ref{Casymp}). In view of
$\varepsilon_+\sim \Mdgp/\MB$ the first term here generates the
first nonanalytic in $\Mdgp$ series of (\ref{efficiency}) and the
second term is responsible for the analytic part because the
hypergeometric function is expandable in Taylor series in
$\varepsilon_+/2\to 0$.\footnote{The $\sigma=m_-/2\MB$ part of the
action originating from the second term of
(\ref{MC_Summary_yRepr_Reduced0}) contributes only to the analytic
part, because $m_-/2\MB\to -1/2$ and the relevant large $\MB$
expansion originates directly from the representation (\ref{5000})
which does not give rise to nonanalytic terms.}

\subsection{The brane effective action in the $O(\CMdim{2})$
approximation and UV divergences} \hspace{\parindent} Substituting
the curvature invariants of (\ref{Abr_Summary}) into
(\ref{TrLnFeff_sum_Cklnu_Aklnu}) we get the lowest orders of the
brane effective action in terms of the curvature invariants
(\ref{zero})-(\ref{two}) listed in Introduction,
    \begin{eqnarray}
     &\frac12\;\mathrm{Tr}\ln{\Feff}
      &\!\!=\,\,
     -\frac12 \frac{M^d}{(4\pi)^{d/2}} \int\limits_{\br}dx\;\sqrt{g}\;
          \MC^{0}_{00}
    \nonumber\\
    &&\quad
     -\frac12\frac{M^{d-1}}{(4\pi)^{d/2}}
     \int\limits_{\br}dx\;\sqrt{g}
          \left\{\left({-}\frac12\MC^0_{01}
          +\frac14\MC^{-1}_{21}\right)\,k
               \;-\; \MC^0_{01} \frac{\Pbr}{2\Mdgp}\,
               \right\}
    \nonumber\\
    &&\quad
     - \frac12\frac{M^{d-2}}{(4\pi)^{d/2}}
     \int\limits_{\br}dx\;\sqrt{g}\;
       \left\{ \fract16\, \MC^{1}_{00}\,
                     \RB
             -\fract{1}{2}\,\MC^0_{11}
                     \PB\right.\nonumber\\
            &&\qquad\qquad\quad
            +\left(-\fract13\,\MC^{1}_{00}\!
                     + \fract14\,\MC^0_{11}\!
                     - \fract18\,\MC^{-1}_{31}\!
               \right) \RB_{nn}
          \nonumber\\
          &&\nonumber\\
          &&\qquad\qquad\quad
            +\; \left( {-}\fract16\,\MC^{1}_{00}\!
                    + \fract{1}{4}\,\MC^0_{11}\!
                    + \fract38\,\MC^{-1}_{31}\!
                    + \fract5{16}\,\MC^{-2}_{51}\!
                    + \fract1{16}\,\MC^{-2}_{42}
              \right)\, k^2_{\za\zb}
          \nonumber\\
          \nonumber\\
          &&\qquad\qquad\quad
            +\left(  \frac16\,\MC^{1}_{00}\!
                    - \frac18\,\MC^0_{11}\!
                    + \frac18\,\MC^0_{02}\!
                    - \frac18\,\MC^{-1}_{22}\!
                    + \frac5{32}\,\MC^{-2}_{51}\!
                    + \frac1{32}\,\MC^{-2}_{42}
              \right)\, k^2
          \nonumber\\
          \nonumber\\
          &&\qquad\qquad\quad
           +\left.\left(\frac12\,\MC^0_{02}\!
       - \fract14\, \MC^{-1}_{22}
       \right)\, \frac{k\,\Pbr}{2\Mdgp}
        +\frac12\,\MC^0_{02}
       \Big(\frac{\Pbr}{2\Mdgp}\Big)^2\,
       \right\}+ O\Big(\CMdim{3}\Big).       \label{TrLnFeff_sum_Cklnu_Aklnu_RESULT}
    \end{eqnarray}
Here the coefficient functions $\MC^j_{kl}$ are given by equations
(\ref{MC_Summary_yRepr_Reduced0})-(\ref{MC_Summary_yRepr_Reduced})
and represent a set of very complicated functions of $\MB$, $\Mdgp$
and $\Mbr$. One can check that for $\Mbr=0$ the $\MC^0_{00}$ term
given by $\MC^0_{00}=\ResHGFp{(0)}{0,\,1}+
(\,\Mp\leftrightarrow\,\Mm\,)$ coincides with the effective
potential calculated for a toy DGP model in \cite{toyDGP}.

More instructive are the ultraviolet divergences of the brane action
which we present here for the four-dimensional case in the
dimensional regularization $d\to 4$. They read
    \begin{eqnarray}
     &\left.\frac12\;\mathrm{Tr}\ln{\Feff}\,\right|^{\,\rm div}
      &\!\!=\,\,
     \frac{1}{32\pi^{2}(4-d)} \int\limits_{\br}dx\sqrt{g}\,
          \Big({-}4\Mdgp^2(\MB^2+2\Mdgp^2-2\Mbr^2)-\Mbr^4\,\Big)
    \nonumber\\
    &&\quad
     +\frac{1}{32\pi^{2}(4-d)}
     \int\limits_{\br}dx\sqrt{g}
          \left(\fract12\Mdgp\big(\MB^2+12\Mdgp^2-3\Mbr^2\big)\,k
               +2\big(\Mbr^2-4\Mdgp^2\big) \Pbr \right)
    \nonumber\\
    &&\quad
     +\frac{1}{32\pi^{2}(4-d)}
     \int\limits_{\br}dx\sqrt{g}
       \left( \fract13(-2\Mdgp^2+\Mbr^2 ) \RB
             +\big( \fract{17}6\Mdgp^2-\fract23\Mbr^2 \big)  \RB_{nn}
             \right.
          \nonumber\\
          &&\qquad\qquad\qquad\qquad\qquad\quad
            +\big( \fract92\Mdgp^2-\fract13\Mbr^2 \big)\, k^2_{\za\zb}
         +\big( {-}2\Mdgp^2+\fract13\Mbr^2 \big)\, k^2\nonumber\\
         &&\qquad\qquad\qquad\qquad\qquad\quad\left.-4\Mdgp^2\PB
         -\frac32\Mdgp\, k\,\Pbr
         -\Pbr^2
       \right)\;\,+\,\;O\big(\CMdim{3}\big). \label{TrLnFeff_sum_Cklnu_Aklnu_d4Pole}
    \end{eqnarray}

This result confirms the general properties of ultraviolet
divergences in any dimension $d$. These divergences are contained in
both series of the expansion (\ref{efficiency}). For an even $d$
they are analytic and polynomial in both $\MB$ and $\Mdgp$, for an
odd $d$ they have a structure $\sqrt{\MB\Mdgp}$ times a finite
polynomial in $\MB$ and $\Mdgp$. Finally, their background
dimensionality is always bounded by $O(\CMdim{d})$. These general
properties follow from the property of the integral
(\ref{Cintegral}) which is UV divergent at the upper limit only for
$d\geq 2j+2l+k$. Therefore, the background dimensionality of the
relevant terms in (\ref{TrLnFeff_sum_Cklnu_Aklnu1}) (with $j=n-c$)
satisfies the bound $k+l+p+2n-2c\leq d+p-l\leq d$, because $p\leq
{\rm max}\{l-2,0\}$. The relevant overall powers of $\Mdgp$ and
$\MB$ in the nonanalytic part of (\ref{TrLnFeff_sum_Cklnu_Aklnu1})
are also positive, because for the same reasons $p+n-c<d/2$ and
$n-c+k+l<d/2$. Finally, in the analytic part of
(\ref{TrLnFeff_sum_Cklnu_Aklnu1}) the overall power of $\MB$ in the
divergent terms is again nonnegative, because $2n-2c+k+2l\leq d$.

\section{The Neumann and Dirichlet limits} \hspace{\parindent}
As we see, the curvature expansion in brane induced gravity models
is essentially more complicated than for pure Dirichlet and Neumann
(Robin) boundary conditions. Even the conformity of our results with
these two limiting cases (corresponding respectively to $\Mdgp\to 0$
and $\Mdgp\to\infty$ in (\ref{1.3})) requires nontrivial
calculations. Here we present these calculations and check the
consistency of the Neumann limit to the $O(\CMdim{2})$ order in the
curvature, and verify the Dirichlet limit to all orders in
$\CMdim{}$.

First we present the known results for a local inverse mass
expansion for pure Dirichlet, $\Phi|_{\br}=0$, and Robin,
$(\partial_n-S)\Phi|_{\br}=0$, boundary conditions (see
\cite{Vassilevich} and references therein). In these two cases
labeled respectively by D and N this expansion for the effective
action in the $(d+1)$-dimensional bulk reads
    \begin{eqnarray}
    &\fracd12\;{\rm Tr}_{D/N}^{(d{+}1)}\ln \FB
    &= -\frac12\int\limits_0^\infty
    \frac{ds}{s}\frac1{(4\pi s)^{(d{+}1)/2}}\, e^{-s M^2}
    \sum\limits_{n=0}^{\infty}s^{n/2} A^{D/N}_n
    \nonumber\\
    &&= -\frac12 \frac{M^d}{(4\pi)^{(d{+}1)/2}}\,
    \sum\limits_{n=0}^{\infty}
    \frac{\GammaF{\frac{-d-1+n}{2}}}{M^{n-1}}\, A^{D/N}_n
    \label{DirichletSDWexpansion}
    \end{eqnarray}
where $A^{D/N}_n$ represent the bulk and boundary integrals of the
relevant Schwinger-DeWitt coefficients. The first four of them for
the Dirichlet case read
    \begin{eqnarray}
    &A^D_0
      &= \int\limits_{\B} d^{d+1}\!X \,\sqrt{G} \;,
    \nonumber\\
    &A^D_1
      &= -\frac{\sqrt{\pi}}2 \int\limits_{\br} d^{d}\!x \,
      \sqrt{g} \;,
    \nonumber\\
    &A^D_2
      &= \int\limits_{\B} d^{d+1}\!X \,\sqrt{G}
      \;\left({-}\PB+\fract16 \RB \right)
        +\int\limits_{\br} d^{d}\!x \,\sqrt{g}
        \;\left({-}\fract{1}{3}\, k\right)\;,
    \nonumber\\
    &A^D_3
      &= \sqrt{\pi} \int\limits_{\br} d^{d}\!x \,\sqrt{g} \;
       \left({+}\fract{1}{2}\PB -
       \fract1{12}\RB + \fract1{24}\,
       \RB_{nn}
        - \fract{7}{192}\, k^2
        +\fract5{96}\, k_{\za\zb}^2\right)\;.        \label{DirichletSDWcoefs}
    \end{eqnarray}

For the Neumann (Robin) case together with the bulk curvature and
the extrinsic curvature of the boundary they involve the coefficient
function $S$ from the Robin boundary condition,
    \begin{eqnarray}
    &A^N_0
      &= \int\limits_{\B} d^{d{+}1}\!X \,
      \sqrt{G} \;,
    \nonumber\\
    &A^N_1
      &= \frac{\sqrt{\pi}}2 \int\limits_{\br} d^{d}\!x
      \,\sqrt{g} \;\;,
    \nonumber\\
    &A^N_2
      &= \int\limits_{\B} d^{d+1}\!X \,\sqrt{G}
      \;\left({-}\PB+\fract16 \RB \right)
        +\int\limits_{\br} d^{d}\!x \,\sqrt{g}
        \;\left({-}\fract{1}{3} k-2S\right)\;,
    \nonumber\\
    &A^N_3
      &= \sqrt{\pi} \int\limits_{\br} d^{d}\!x \,\sqrt{g} \;
       \left({-}\fract{1}{2}\PB + \fract1{12}\RB
       - \fract1{24}\,
      \RB_{n n}
       \right.\nonumber\\
       &&\qquad\qquad\qquad\quad
        \left.
        +\fract{13}{192}\, k^2
        +\fract1{96}\, k_{\za\zb}^2
        + \fract12\, k S
        +S^2\right)\;.    \label{NeumannSDWcoefs}
    \end{eqnarray}

  \subsection{The Neumann (Robin) limit.}
  \hspace{\parindent}
The Robin limit of the boundary condition (\ref{1.3}) corresponds to
$\Mdgp\to+\infty$ with $\MB/\Mdgp=0$, $\Mbr/\Mdgp=0$ and the finite
limiting value of $\Pbr/2\Mdgp\to S$. This implies the following
limits for the auxiliary mass parameters
    \begin{eqnarray}
     \Mp\to 0\,,\qquad
     \Mm\to -\infty\,,\qquad
     \hdM\to +\infty\,,\qquad
     \fract{\Mdgp}{\hdM}\to 1\,.
     \label{AuxMassesNeumannLimit}
    \end{eqnarray}
Therefore the contribution of $\sigma=m_-/2\MB\to-\infty$ terms in
the coefficients $\MC^j_{kl}(\MB,\Mpm)$ given by
(\ref{MC_Summary_yRepr_Reduced0})-(\ref{ResHGF_def}) vanishes in
virtue of the asymptotic behavior of the hypergeometric function
$\HGF{a,\,b}{c}{z}$ at $z\to-\infty$. Moreover, only the functions
$\ResHGFp{(n)}{a,\,b}$ with a special combination of the
hypergeometric function's indices (\ref{5000}) arise, and due to the
relation 
    \begin{eqnarray}
    &&\HGF{a,\,b}{\frac{a}2{+}\frac{b}2{+}\frac12}{\frac12}
       =  \frac{\GammaF{\frac12}\GammaF{\frac{a}2{+}
       \frac{b}2{+}\frac12}}{\GammaF{\frac{a}2{+}
       \frac12}\GammaF{\frac{b}2{+}\frac12}}   \nonumber 
    \end{eqnarray}
they equal
    \begin{eqnarray}
     &\ResHGFp{(n)}{a,\,b}
      &= \frac{\GammaF{\fract12}}{n\,!}
          \frac{\GammaF{a{+}n}\GammaF{b{+}n}}
          {\GammaF{\fract{a}2{+}\fract{n}2{+}
          \fract12}\GammaF{\fract{b}2{+}
          \fract{n}2{+}\fract12}}=\frac{2^{a{+}b{+}2n{-}2}}{n\,!\,\GammaF{\fract12}}
          \GammaF{\fract{a}2{+}\fract{n}2}\GammaF{\fract{b}2{+}\fract{n}2}\;,
    \nonumber
    \end{eqnarray}
where we used the gamma function identity $\GammaF{2z}\GammaF{1/2} =
2^{2z{-}1}\, \GammaF{z}\, \GammaF{z{+}1/2}$. As a result the
coefficient functions $\MC^j_{kl}$ reduce to
    \begin{eqnarray}
     &&\MC^j_{00}
     \;=\; \frac12 \GammaF{j-d/2},\,\,\,\,
     \MC^j_{kl}
     \;=\;
     \frac{\GammaF{j{+}\frac{k{+}l-d}2}}
     {\GammaF{\frac{k{+}l}2}},\,k+l>0\;,         \label{MC_klnu_NeumannLimit}
    \end{eqnarray}
and the expansion (\ref{TrLnFeff_sum_Cklnu_Aklnu_RESULT}) for the
case of the Neumann boundary conditions takes the form
    \begin{eqnarray}
     &\frac12\;\mathrm{Tr}\ln{\Feff}
      &= - \frac12\frac{\MB^d}{(4\pi)^{d/2}}\,
      \sum\limits_{n=0}^\infty
      \frac{\GammaF{{-}\frac{d}2{+}\frac{n}2}}{\MB^n}
     \, \Abr_n \;,                     \label{TrLnFeff_sum_Neumann}
    \end{eqnarray}
where the first three coefficients equal
    \begin{eqnarray}
     &\Abr_0
      &=  \frac12 \brint \;\;,
      \nonumber\\
     &\Abr_1
      &= -\fract1{\sqrt{\pi}}
      \brint \;\fract{\Pbr}{2\Mdgp}\;,
      \nonumber\\
     &\Abr_2
      & = \brint \;\left(
      \fract1{12} \RB - \fract1{24} \RB_{n n }
      -\fract1{48}k_{\za\zb}^2+\fract5{96} k^2
      \right.\nonumber\\
       &&\qquad\qquad\qquad\quad
        \left.
      -\fract12 \PB +\fract14 k \left(\fract{\Pbr}{2\Mdgp}\right)
      +\fract12 \left(\fract{\Pbr}{2\Mdgp}\right)^2 \right) \;.
    \end{eqnarray}

The consistency of this result with the Robin case is based on the
duality relation (\ref{Neumann-to-Dirichlet_Reduction}) which
implies the following identities for the coefficients of the
Schwinger-DeWitt expansion (\ref{DirichletSDWexpansion})
    \begin{eqnarray}
     \Abr_n
      =\frac1{\sqrt{4\pi}}\, \big( A^N_{n{+}1} - A^D_{n{+}1}\big).
    \end{eqnarray}
These identities can be directly checked for
(\ref{DirichletSDWcoefs})-(\ref{NeumannSDWcoefs}) with
$S=\Pbr/2\Mdgp$ for $n=0,1,2$.

\subsection{The Dirichlet case}
\hspace{\parindent} The Dirichlet limit formally corresponds to
$\Mdgp\to 0$ when $\Mdgp/\MB\to0$, $\Mdgp/\Mbr\to 0$ and
    \begin{eqnarray}
     \Mpm\to \pm\sqrt{\MB^2{-}\Mbr^2}\,.
     \label{AuxMassesDirichletLimit}
    \end{eqnarray}
This limit cannot however be directly taken in the expansion
(\ref{efficiency}) because of its obvious non-analyticity at
$\Mdgp=0$. The reason is that for small $\Mdgp$ the behavior
(\ref{Casymp}) of $C^j_{kl}$ used in the original curvature
expansion series (\ref{TrLnFeff_sum_Cklnu_Aklnu}) no longer applies.
Indeed, when $\Mdgp \MB<\mu^2$ the parameter $\varepsilon_+$ in the
integral (\ref{Cintegral}) has another asymptotics $\mu^2/2M^2$
independent of $\Mdgp$, and $C^j_{kl}\sim m^l$ for $\Mdgp\to 0$.
Therefore, in this range of $\Mdgp$ the DGP scale arises in
(\ref{TrLnFeff_sum_Cklnu_Aklnu}) only in positive powers $l-p>0$,
because $p\leq l-2$ --- the curvature expansion takes the form
qualitatively different from (\ref{efficiency}). The only
nonvanishing term in the sum over $k$ and $l$ is the one with
$k=l=0$ (cf. equation (\ref{Wk0})). The relevant coefficients
(\ref{MC_Summary_yRepr_Reduced}) are
    \begin{eqnarray}
     \MC^j_{00}(\MB,\Mpm)=
     \frac1{2} \int\limits_0^\infty \frac{d\pt}{\pt}
     \pt^\nu \big(w(-\Mp\!\sqrt{\pt})
      +w(-\Mm\!\sqrt{\pt})\big)\: \,e^{-\pt \MB^2}
      =  \frac{\GammaF{j-d/2}}{\Mbr^{2j-d}} \;,
    \end{eqnarray}
because in the Dirichlet limit $\Mp^2=\Mm^2=\MB^2{-}\Mbr^2$ and the
error functions in the definition of the proper time weight
$w(-\Mp\!\sqrt{\pt})$ satisfy
$\erfc{(-\Mpm\!\sqrt{\pt})}=1\pm\erf{(\sqrt{\MB^2{-}\Mbr^2}\,\sqrt{\pt})}$.
These are the prefactors of the integrated Schwinger-DeWitt
coefficients $A_{2j}=\int d^dx\,a_j(x,x)$ in the expansion of the
$d$-dimensional effective action for the massive operator
(\ref{hFbr_def})
    \begin{eqnarray}
     \frac12\,
     \mathrm{Tr}\,\ln {\Fbr}(\nabla)
     =\frac12\,
     \mathrm{Tr}\,\ln{(-\brBox{+}\Mbr^2{+}\Pbr)}
      = - \frac12\,\frac{\Mbr^d}{(4\pi)^{d/2}}
      \sum\limits_{j=0}^\infty
      \frac{\GammaF{j-d/2}}{\Mbr^{2j}}    \label{10000}
     \, A_{2j} \;.
    \end{eqnarray}
This can be seen already at the level of $\HW$-expansion
(\ref{1000}). In the limit $\Mdgp\to0$ all perturbation terms in
(\ref{1000}) with $1/\Feffo=2\Mdgp/(\Mbr^2-\brBox)$ vanish except
the contribution of $U_1/\Feffo=-\Pbr/(\Mbr^2-\brBox)$, which
complements the first term of this expansion to theone-loop action
of the full brane operator with a potential (\ref{10000}) (we of
course disregard the volume divergent ${\rm Tr}\,\ln{2\Mdgp}$ part
canceled by the local measure).

Thus under the $\Mdgp\to0$ limit the Neumann-to-Dirichlet reduction
(\ref{Neumann-to-Dirichlet_Reduction}) leads to
    \begin{eqnarray}
     \lim\limits_{\Mdgp\to \,0} \frac12\;
     {\rm Tr}_N^{(d+1)}\ln \FB
     =\frac12\;{\rm Tr}_D^{(d+1)}\ln \FB
     + \frac12{\rm Tr}
     \ln{\Fbr}\;.   \label{NtoD_Reduction_Mdgp_to_0_Limit}
    \end{eqnarray}

The second term here originates from the definition of the
generalized Neumann vs Dirichlet functional determinants of the
operator $\FB$
    \begin{eqnarray}
     &&\exp\Big(-\frac12\;
      {\rm Tr}_N^{(d+1)}\ln \FB\Big)=\int
     D\varPhi\,\exp\Big(-S_\B[\,\varPhi\,]-\frac12\brint
     \varphi\,\Fbr(\nabla)\,\varphi\Big),\\
     &&\exp\Big(-\frac12\;
      {\rm Tr}_D^{(d+1)}\ln \FB\Big)=\int
     D\varPhi\,\exp\Big(-S_\B[\,\varPhi\,]\Big)\,\delta\big(\,\varphi(x)\,\big),
    \end{eqnarray}
where $S_\B[\,\varPhi\,]$ is the bulk part of the action
(\ref{QuadraticAction}) and the expression for the delta function of
$\varphi(x)=\varPhi(X)\,|_\B$ (regularized by $\Mdgp\to 0$) reads as
    \begin{eqnarray}
     \delta\big(\,\varphi(x)\,\big)=\lim\limits_{\Mdgp\to \,0}\;
     \Big({\rm Det}\,\Fbr\Big)^{1/2}
      \exp\Big(-\frac12\brint
     \varphi\Fbr(\nabla)\varphi\Big).
    \end{eqnarray}
The preexponential factor here, $\big({\rm Det}\,
\Fbr\big)^{1/2}=\exp\big({\rm Tr}\,\ln \Fbr/2\big)$, gives rise to
the last term in (\ref{NtoD_Reduction_Mdgp_to_0_Limit}) and, thus,
confirms the consistency of the Dirichlet limit.

\section{Conclusions}
  \hspace{\parindent}
Thus we have constructed the covariant curvature expansion in
massive brane induced gravity models, found its peculiar structure
(\ref{efficiency0}) nonanalytic in the DGP scale and derived a
nontrivial cutoff (\ref{cutoff}) of this general expansion. Finally,
we calculated several lowest orders of this expansion for a quantum
scalar field in a curved bulk spacetime with a kinetic term on the
brane to a quadratic order in background dimensionality and found
its ultraviolet divergences for the case of a 4-dimensional brane.

These results might find important applications. Although a
comparison of our massive model with the massless DGP model of
\cite{DGP} is not straightforward, we can observe a common feature
in their cutoff properties. In both theories their cutoff
(\ref{cutoff}) is different from the bulk one $\MB$ and is modified
by the DGP scale $\Mdgp$. For the tree-level DGP model with the
Planck mass $\MB=M_P$, playing the role of the bulk cutoff, this
cutoff equals $M_{\rm cutoff}=(m^2 M)^{1/3}$ \cite{scale}. With
$\Mdgp$ identified with the cosmological horizon scale, this is
about $(1000 \rm km)^{-1}$ which is much below the submillimeter
scale capable of featuring the infrared modifications of the
Einstein theory \cite{Gundlach}. As we see, the situation with the
local expansion for the {\em quantum} action is much better --- the
cutoff (\ref{cutoff}) is a geometric average of $\MB$ and $\Mdgp$,
which is much higher,
    \begin{eqnarray}
    (m M)^{1/2}\gg(m^2 M)^{1/3},
    \end{eqnarray}
and comprises $(0.1 \rm mm)^{-1}$. This supports the conjecture that
the replacement of the weak field perturbation theory by a
derivative expansion \cite{Dvali}, as is the case of the local
Schwinger-DeWitt series, might improve the range of validity of the
calculational scheme.

Obviously, the Schwinger-DeWitt technique in brane induced gravity
models turns out to be much more complicated than in models without
spacetime boundaries or in case of boundaries with local Dirichlet
and Neumann boundary conditions. It does not reduce to a simple
bookkeeping of local surface terms like the one reviewed in
\cite{Vassilevich}. Nevertheless it looks complete and
self-contained, because it provides in a systematic way a manifestly
covariant calculational procedure for a wide class of boundary
conditions including tangential derivatives (in fact of any order).
On the other hand, the calculational strategy of the above type
requires a further extension, because there is still a large set of
issues and possible generalizations to be resolved in concrete
problems.

One important generalization is a physically most interesting limit
of a vanishing bulk mass $M^2$, whose rigorous treatment should
justify a qualitative comparison of the above type for the cutoff
scales in our expansion (\ref{efficiency}) and the weak field
expansion of the DGP model. The local curvature expansion is perfect
and nonsingular for nonvanishing $M^2$ and is applicable within its
cutoff scale (\ref{cutoff}). However, for $M^2\to 0$ it obviously
breaks down, because the proper time integrals start diverging at
the upper limit and all UV finite terms of (\ref{efficiency0}) blow
up. These infrared divergences can be avoided by a nonlocal
curvature expansion of the heat kernel of \cite{CPT}. Up to the
cubic order in curvatures this expansion explicitly exists for ${\rm
Tr}\,e^{s\Box}$ \cite{CPT3}, but for the structure involving a local
differential operator ${\rm Tr}\,W(\nabla) e^{s\Box}$ it still has
to be developed.

Another important generalization is the extension of these
calculations to the cases when already the lowest order
approximation involves a curved spacetime background (i.e. dS or AdS
bulk geometry, deSitter rather than flat brane, etc.). The success
of the above technique is obviously based on the exact knowledge of
the $y$-dependence in the lowest order Green's function in the bulk
and the possibility to perform exactly (or asymptotically for large
$M^2$) the integration over $y$. All these generalizations and open
issues are currently under study.

To summarize, we developed a new scheme of calculating quantum
effective action for the braneworld DGP-type system in curved
spacetime. This scheme gives a systematic curvature expansion by
means of a manifestly covariant technique. Combined with the method
of fixing the background covariant gauge for diffeomorphism
invariance developed in \cite{gospel,qeastbg} this gives the
universal background field method of the Schwinger-DeWitt type in
gravitational brane systems.

\section*{Acknowledgements}
  \hspace{\parindent}
A.B. is grateful for hospitality of the Laboratory MPT CNRS-UMR 6083
of the University of Tours, where a part of this work has been done.
The work of A.B. was supported by the Russian Foundation for Basic
Research under the grant No 08-01-00737. The work of D.N. was
supported by the RFBR grant No 08-02-00725 and a research grant of Russian
Science Support Foundation. This work was also supported by the LSS
grant No 1615.2008.2.


\end{document}